\documentclass[12pt]{article}
\usepackage{hyperref}
\usepackage{amsmath}
\usepackage{amsfonts}
\usepackage{amssymb}
\usepackage{tensor}
\usepackage{graphicx}
\usepackage{color} %for todo's
\usepackage{slashed} %\shashed{p}
\usepackage{subfig}

\usepackage{tikz}
\usetikzlibrary{arrows}

\numberwithin{equation}{section} % for eq. numbered like (3.1)
\newcommand{\lsim}{\mbox{\raisebox{-.9ex}{~$\stackrel{\mbox{$<$}}{\sim}$~}}}
\newcommand{\gsim}{\mbox{\raisebox{-.9ex}{~$\stackrel{\mbox{$>$}}{\sim}$~}}}
\newcommand{\R}{\mathbb{R}}
\newcommand{\La}{\mathcal{L}}

\newcommand\p{\partial}

\newcommand{\be}{\begin{equation}}
\newcommand{\ee}{\end{equation}}

\begin{document}

\title{Randall-Sundrum Models and Holography}
\author{Jeremy Price \\ \quad \\  \emph{Department of Physics, University of Washington, Seattle, WA 98195, USA}}
\date{}

\maketitle

\begin{center} {\bfseries Abstract} \end{center}

This Master's thesis will introduce the Randall-Sundrum RSI model and overview its general features and basic phenomenology, with the goal of understanding its 5-dimensional geometric structure in terms of its 4-dimensional AdS/CFT dual theory. We discuss mode expansions, boundary terms, field content, and supersymmetry, leading up to a discussion of the AdS/CFT correspondence in RSI models. Low-energy 4d Yukawa hierarchies and gauge couplings are understood in terms of the bulk 5d theory. A few simple Standard Model and MSSM-like constructions are considered, and their 4d dual theories are discussed.  One new result is presented: the 5d bulk interpretation of the Nelson-Strassler mechanism of generating Yukawa hierarchies through renormalization group flow. 

%\nocite{*} %use to include all references to find uncited ones
\newpage

\tableofcontents

\section{Introduction}

In the past ten years, theories with extra dimensions have received an enormous amount attention from model builders.  Although the basic ideas of extra dimensions have been around since the theories of Kaluza and Klein in the 1920s, they had serious problems, such as the lack of chiral fermions and the existence of new light particles, that were not understood until recently. Much of this recent attention has been through the construction of Randall-Sundrum and ADD models.  

In the 1980s Witten showed \cite{Witten:1983ux} that in Kaluza-Klein theories on smooth coset spaces, an index theorem prevents the existence of chiral fermions. To circumvent this theorem, models on singular spaces such as orbifolds or coniflds may be constructed, where boundary conditions at singular points ensure that chiral fermions are allowed.  This did not solve the problem of light modes, which generally appeared to be inconsistent with observed interactions and cosmological observations \cite{PerezLorenzana:2005iv}.

Later, in the 1990s it was discovered that certain string theory backgrounds could admit solitonic `domain-wall' like solutions, called branes. The low-energy effective theories associated with branes placed at singular points were found to be examples much like the old Kaluza-Klein theories on singular backgrounds allowing for chiral fermions.  With these models, theories could be constructed with particle content identical to the Standard Model's.

Although naive estimates of the size of extra dimensions, due to the inclusion of gravity, are of order the Planck length, models were soon proposed where the size of the extra dimensions could be much larger than that--possibly even infinite--and still remain undetected.  Phenomenologically reasonable models of extra dimensions have recently been proposed by Horava and Witten \cite{Horava:1996ma}; Randall and Sundrum \cite{Randall:1999ee,Randall:1999vf}; Arkani-Hamed, Dimopoulos, and Dvali \cite{ArkaniHamed:1998rs}; and others.

Independently from these constructions, in 1998, Maldacena \cite{Maldacena:1997re} discovered an unexpected duality between certain types of string theories on particular backgrounds and strongly coupled, large $N$ SCFTs. This duality allowed the models with AdS backgrounds, such as Randall-Sundrum, to be investigated in an entirely new way. Although explicit embeddings of Randall-Sundrum models into string theory are not known, the correspondence can still be used. In these examples, light fields such as the radion need not be a problem, and may be consistent with known measurements.

The purpose of these notes is to review Randall-Sundrum's `RSI' construction of a 5-dimensional theory in an interval of AdS space and investigate its basic features, then, to introduce supersymmetry, and use the AdS/CFT correspondence to investigate features of the 4d dual theories in terms of the 5d bulk.

First, we will outline the basic construction of warped models, discuss the basic fields, their expansions, and boundary conditions. Then a few simple connections to the Standard Model will be discussed, where we will see how this model solves the hierarchy problem and addresses flavor universality. Supersymmetry will be introduced, and its consequences for simple 5d MSSM-like models will be considered. Then AdS/CFT will be overviewed, and we will translate our setup into a 4d dual language, where 5d geometric properties will show up in unexpected ways. Finally, one minor new result will given: the bulk interpretation of \cite{Nelson:2000sn} will be discussed explicitly, which has not been done in the literature, although 5d models which have similar 4d duals have been constructed \cite{Benini:2009ff, Klebanov:2000hb}.
\section{A Flat Extra Dimension}

We will start by considering a 5-dimensional spacetime with coordinates $x^M$, $M=(\mu,5)$, $\mu=0,1,2,3$. The 5-coordinate will usually be denoted by $y=x^5$. Because simple models of large extra dimensions are ruled out by direct observation we will assume that the 5-coordinate corresponds to a coordinate on a compact one dimensional space which is `small' enough that it has not been directly detected. The arguments for the smallness of this extra dimension go back to the old Kaluza-Klein models of the 1930s and essentially require the characteristic size of the extra dimension to be smaller than the smallest distances gravity has been tested to, since modifications to the gravitational potential due to the extra compact dimensions can be parameterized by \cite{Randall:1999vf}
\be
	V(r) = V_\text{Newton}(r) (1 + a e^{-\alpha r}/r).
\ee

The simplest choice for this extra dimension is the circle, and so we assume our 5-dimensional space is given by the product space $\R^{3,1}\times S^1$.  However, as mentioned in the introduction, it is a theorem of Witten's that theories with chiral fermions cannot exist in higher dimensional smooth manifolds, so for a realistic phenomenology this is not acceptable \cite{Sundrum:2005jf}.  

A simple way around this objection is to quotient the circle by the natural $\mathbb{Z}_2$ action identifying antipodal points. The new space obtained is a singular space called an orbifold, which in this case is isomorphic to a line segment with singular endpoints\footnote{It is a well-known theorem that fixed points of a group action correspond to singular points when a space is quotiented by the action.  The endpoints are fixed points under the group action: clearly $0\rightarrow -0$ is a fixed point, and $\pi R$ is a fixed point because $\pi R \rightarrow -\pi R \cong -\pi R + 2\pi R = \pi R$.}.  All fields will be required to have a definite parity under this $\mathbb{Z}_2$ action.  By convention we choose the circle to have radius $R$, which, up to translation, fixes the singular endpoints to be at $0$ and $\pi R$. 

To construct our theory, at each orbifold singularity we place a 3-brane which extends infinitely in the $\R^{3,1}$ direction. For reasons we will soon see, the brane at $y=0$ will be called the UV brane, and the brane at $y=\pi R$ the IR brane.

As we have stated it, the theory so far is 5d gravity (since we have only specified coordinate fields) on the flat space $\R^{3,1}\times S^1/\mathbb{Z}_2$. From here, one can investigate the consequences of 5d gravity, or add bulk fields to the 5d space \cite{Csaki:2004ay}. What we will want to do next, however, is to warp the extra dimension by adding a bulk cosmological constant. 

\section{A Warped Extra Dimension}

We can take the flat metric on our orbifolded theory defined above and add a bulk (5d) negative cosmological constant $\Lambda < 0$. The metric for this space, which is just AdS times an interval, is
\be
	ds^2 = e^{-2ky}\eta_{\mu\nu} dx^\mu dx^\nu + dy^2 = g_{MN} dx^M dx^N
\ee
where $k$ is the curvature radius of the embedded AdS space, and throughout we will use the mostly plus metric. This is the Randall-Sundrum model RSI (confusingly, the RSI model has two branes, and there is also the RSII model, which is the same, except that it has one brane) \cite{Randall:1999ee, Randall:1999vf}.

This model can be thought of as a slice of AdS$_5$, with an effective UV brane at one end, and an IR brane at the `bottom' of the AdS throat. This is reminiscent of the Klebanov-Strassler solution in AdS$_5 \times S^5$ where fluxes close off the IR end of the geometry. In this picture UV brane should be thought of as some effective brane that is really some complicated Calabi-Yau whose degrees of freedom are integrated out, and associated `fundamental' fields that come from the structure of the Calabi-Yau are simply placed by hand on the effective UV brane. This picture will be useful to us later when we try to understand the full 4d dual theory to these 5d theories using the AdS/CFT correspondence. 

\subsection{Scales}
\label{scales}

Before we start looking at the details of this model, let's try to look at the consequences of warping the 5th dimension though the introduction of a bulk cosmological constant. Let us imagine for simplicity that the entire Standard Model is confined to the IR brane, located at $y=\pi R$. We can write a simple action of a scalar field $H$ in 5d, that we can imagine to be a simple model of a Higgs.
\be
	S_H = -\int d^5x \sqrt{-g} \left( \p_\mu H^\dagger \p^\mu H - M_5^2 \left|H\right|^2 + \lambda \left|H\right|^4 \right) \delta(y-\pi R)
\ee
where $M_5$ is the bulk mass of the Higgs. Doing the $y$ integral easily gives us,
\be
	S_H = - \int d^4x \left( e^{-2\pi k R}\p_\mu H^\dagger \p^\mu H - M_5^2 e^{-4\pi k R} \left|H\right|^2 + \lambda e^{-4\pi k R}\left|H\right|^4 \right).
\ee
To understand this we should rescale $H \rightarrow e^{-\pi k R} H$ so that it has a canonical kinetic term.
\be
	S_H = -\int d^4x \left( \p_\mu H^\dagger \p^\mu H - M_5^2e^{-2\pi k R} \left|H\right|^2 + \lambda \left|H\right|^4 \right).
\ee
We see that an observer living on the IR brane sees the Higgs to have a redshifted mass which is exponentially smaller than the bulk mass.  If we assume that the Higgs had a natural (order one in natural units) mass in the 5d theory, then $M_5 \approx M_\text{Planck}$, so that the 4d Higgs mass is,
\be
	\label{hierarchies:higgs-mass}
	m_\text{Higgs} = M_5 e^{-\pi kR}.
\ee
If we assume the curvature of AdS is also order one, so that $k\approx M_\text{Planck}$, and assume that $R$ is of order $10$, the 4d Higgs mass is
\be
	m_\text{Higgs} \approx \text{GeV - TeV}.
\ee
This is a very nice result. it reduces the hierarchy between the Planck and Higgs masses from 16 orders of magnitude to a tuning of about one order of magnitude. So warping the extra dimension looks like a promising way to solve the hierarchy between the weak and Planck scales by reducing an `exponential' hierarchy to a `linear' one.

It is easy to see that any mass scale associated with the IR end of the geometry will be redshifted. One can hope that this can be used to explain a variety of features, such as generating a fermion mass hierarchy starting with order one 5d masses, or exponentially suppressing higher dimension operators associated with FCNCs and processes like proton decay.  Often the IR scale is taken to be $\sim$TeV, and is responsible for the generation of TeV scale soft masses, with the Higgs living on the IR brane.  This is not needed in order to explain hierarchies or exponential suppression, however, and as we will see later in the supersymmetric case the IR scale generally corresponds to a scale very different than the weak scale.

\subsection{Radius Stabilization}
\label{Goldberger-Wise}
This hierarchy-generating process relies on an important mechanism which we haven't yet discussed. It is clear that the exponential warp factor in the metric was the important part of this discussion, but it seems that we can eliminate it through a change of the $dx$ coordinates,
\be
	ds^2 = e^{-2\sigma}\eta_{\mu\nu} dx^\mu dx^\nu + dy^2 \rightarrow \eta_{\mu\nu} d\tilde{x}^\mu d\tilde{x}^\nu + dy.
\ee
To understand how this problem is resolved, recall the way 5d tensors transform in terms of 4d components (as was seen in the original Kaluza-Klein theories' decomposition of the 5d metric into gravitational and electromagnetic pieces),
\begin{equation}
\nonumber
	g^{MN} = 
	\left(\begin{array}{ccc|c}
		& & & \\
		 & g^{\mu\nu}  & & A^{\mu} \\
		 & & & \\
		\hline
		 & A^{\mu} &   & \phi \\
	\end{array}\right).
\end{equation}
In other words, a 5d tensor can be decomposed as $(\text{4d tensor}) \otimes (\text{4d vector}) \otimes (\text{scalar})$.  This means that in the 5d pure gravity theory, fluctuations of the 5d metric will cause corresponding fluctuations associated with the massless 4d graviton, a massless 4d vector, and a massless scalar. 

The 5d gravity solutions do not involve the radius $R$, so if it were treated as a dynamical field, there would be no associated potential, meaning the corresponding field is a massless scalar field. Because the only scalar in the 5d theory is $\phi$, the $55$ component of the metric, its vacuum expectation value must determine the radius.  In order for a specific radius to be chosen we must find a way to generate a potential for $\phi$, which will cause it to develop an expectation value, and turn it into a massive field.

The simplest way to describe how this can happen is with the Goldberger-Wise mechanism \cite{Goldberger:1999uk}. The details won't be important to our discussion below, so we will only summarize the results.

We can introduce potentials for the radion field $\phi$ on the UV and IR branes,
\be
	S_\phi = - \int d^4 x \sqrt{-g_i} \lambda_i (\phi^2-v_i^2)^2
\ee
where $i=$ UV, IR. These terms cause $\phi$ to develop $y$-dependent vacuum expectation values at the minimum of a potential $V_\phi$. The minimum of this potential can be found to be at
\be
	kR = \frac{4}{\pi} \frac{k^2}{m^2} \ln\frac{v_\text{UV}}{v_\text{IR}}.
\ee
This means we can get a $kR \sim 1-10$ with $\frac{k^2}{m^2} \sim 1-10$ and the couplings of order one, or $\frac{k^2}{m^2} \lsim 10$ and $\ln\frac{v_\text{UV}}{v_\text{IR}} \gsim 1$. So no large fine-tunings are required, and a $kR$ of the right size can be naturally obtained. 

\section{Bulk Fields Without SUSY}

Now we will write down the gravitational action for our 5 dimensional space. For a review of gravity oriented towards particle physics, including supergravity and higher dimensions, see \cite{Ortin:2004ms}.  The action for pure 5 dimensional gravity in our setup, with domain walls at $y=0,\pi R$ is given by
\be
	S = S_\text{5d bulk} + S_\text{UV brane} + S_\text{IR brane}
\ee
\be
	\label{bulk:gravity}
	S_\text{5d bulk} = -\int d^5x \sqrt{-g} \left[ \frac{M_5^3}{2} \mathcal{R} +  \Lambda \right]
\ee
\be
	S_\text{UV+IR} = -\int d^5x \sqrt{-g}\left[ \delta(y)(\Lambda_\text{UV} + \La_\text{UV}) + \delta(y-\pi R) (\Lambda_\text{IR} + \La_\text{IR}) \right].
\ee
Here, $M_5$ is the 5d Planck mass, $\mathcal{R}$ is the 5d scalar curvature, $\Lambda$ is the 5d cosmological constant, $\Lambda_{\text{IR/UV}}$ are the cosmological constants on the branes, and $\La_\text{UV/IR}$ are the Lagrangians on the branes. 

Formally, we treat the $y$ coordinate as running from $-\pi R$ to $\pi R$, but because the $\mathbb{Z}_2$ action identifies $y$ with $-y$, all $y$ values can be treated as positive and ranging between the group action's fixed points at $0$ and $\pi R$. Some care about which convention is used is needed when dealing with normalizations and boundary conditions.

The solution for the 5d metric is given by 
\be
	ds^2 = g_{MN} dx^M dx^N =  e^{-2\sigma} \eta_{\mu\nu} dx^\mu dx^\nu + dy^2
\ee
where we have defined the dimensionless $\sigma = k|y|$, and as before $k$ is the AdS curvature radius.  For reference,
\be
	g = \text{det}(g^{MN}) = - (e^{-2\sigma})^4, \quad \sqrt{-g} = e^{-4\sigma}
\ee
\be
	g^{MN} \p_M = e^{+2\sigma} \eta^{MN} \p_M = \p_N.
\ee

In order for the metric given above to be a consistent solution to the 5d action, the Einstein field equations enforce that the bulk and boundary cosmological constants be related \cite{Randall:1999ee}.
\be
	\Lambda = - 6 M_5^3 k^2
\ee
\be
	\Lambda_\text{IR} = -\Lambda_\text{UV} = -\frac{\Lambda}{k}.
\ee
We can define the effective mass scale on the IR brane to be the redshifted UV mass $M_P$,
\be
	M_\text{IR} = M_P e^{-\pi k R}
\ee
as we saw in section \ref{hierarchies:higgs-mass}.  The effective mass scale on the UV brane is the four dimensional Planck mass $M_P$, defined through relating the 4d brane and 5d bulk actions, and is given by \cite{Randall:1999ee}
\be
	M_P^2 = \frac{M_5^3}{k}(1-e^{-2\pi k R}).
\ee
For the rough values used earlier for $k$ and $R$, this would give $M_5 \approx M_P$, since the exponential would be small and $k\approx M_5$.   This justifies out earlier terminology of ``UV'' and ``IR'' branes, since the UV brane is at the unredshifted end and the IR brane is at the redshifted end.

We also note for future use that the derivatives of $\sigma$ are given by
\be
	\sigma' = k \epsilon(y)
\ee
\be
	\sigma'' = 2k(\delta(y) - \delta(y-\pi R))
\ee
where $\epsilon$ is the step function that is $+1$ for positive $y$ and $-1$ for negative $y$.

Now that we have the basic setup, we can start looking at the different fields which can live in our model.

\subsection{Mode Expansions}
\label{bulkfields:expansions}

We will want to add fields to the bulk, separate variables and expand them in Kaluza-Klein modes so we can treat the $x$ and $y$ dependences separately.  Scalar, spinor, vector, and tensor fields can be all treated in detail at once \cite{Gherghetta:2000qt}, but we will overview the generalities and then treat each case individually.  Schematically we want to expand a generic bulk field $\Phi$ as
\be
	\Phi(x^\mu,y) = \sum_{n=0}^\infty \Phi^{(n)}(x) f_n(y) 
\ee
where $\Phi$ are eigenfunctions of some operator whose eigenvalues are the Kaluza-Klein masses $m_n$. Wave function profiles $f_n$ satisfy an orthonormality condition and have differential equations which we must solve to understand the bulk behavior of the field, and determine the masses $m_n$.

Boundary conditions of the fields at the UV and IR branes are encoded in the wave function profiles $f_n$.  There are two cases to consider, when the field is even or odd under the $\mathbb{Z}_2$ symmetry.

If the field $\Phi$ is even, then under the group action, $f_n(y) \rightarrow f_n(|y|)$. This means the derivative $f'_n$ at the boundary is a step function which as we will show later is proportional to the field's bulk mass,
\be
	\left.\frac{df_n}{dy}\right|_\text{UV/IR} = \left. r \sigma' f_n \right|_\text{UV/IR}
\ee
where $r$ is the mass parameter for the specific field, which will be 0 or the $b$ or $\pm c$ defined below. 

If the field $\Phi$ is odd, then $f_n(y)\rightarrow \epsilon(y) f_n(|y|)$. By continuity, this means
\be
	\left. f_n \right|_\text{UV/IR} = 0.
\ee

So even/odd parity of the 5d field determines if we use Neumann or Dirichlet boundary conditions, and tells us what the exact boundary conditions are in terms of the masses. These boundary conditions correspond to adding boundary actions on the UV and IR branes, so this can be implemented by specifying boundary conditions, or equivalently requiring the variation of the boundary action to vanish. A detailed discussion of boundary kinetic terms in Randall-Sundrum models can be found in \cite{Davoudiasl:2002ua}. In the case of fermions, this is discussed in \cite{Contino:2004vy}. 

\subsection{Scalar Fields}

Let $\phi$ be a bulk complex scalar field with bulk action
\be
	\label{scalar:action}
	S_\phi = -\int d^5x \sqrt{-g} \left( \left| \p_M \phi \right|^2 + m_\phi^2 \left|\phi\right|^2 \right)
\ee
and with boundary action
\be
	\label{scalar:boundaryaction}
	S_\text{IR/UV} = -\int d^5x \sqrt{-g} (2m_b\left)( \delta(y)-\delta(y-\pi R) \right) |\phi|^2.
\ee
As we have mentioned, generically, fields in the bulk will allow boundary actions, unless forbidden by symmetries, and these boundary actions can be seen as being required by the variational principle in order to cancel boundary variations. Because these boundary terms are related to bulk variations through the variational principle, we should expect that any boundary mass term is generally determined by the bulk masses. 

Because $k$ is a natural scale in 5 dimensions with units of mass, let us define a parameter, $a$, called the `bulk mass parameter' for $\phi$ by $m^2_\phi = a k^2$, and a `boundary mass parameter' $b$ by $m_b = b k$.  So we may absorb the boundary term into the bulk by writing the bulk mass as $m_\phi^2 = ak^2 + 2bk (\delta(y)-\delta(y-\pi R))$.  The equations of motion are given by
\be
	\frac{1}{\sqrt{-g}} \p_M \left[ \sqrt{-g} g^{MN} \p_N \phi \right] - m^2_\phi = 0.
\ee
To solve this, we can use separation of variables and expand in Kaluza-Klein modes,
\be
	\phi(x^M) = \phi(x^\mu,y) = \sum_{n=0}^\infty \phi^{(n)}(x^\mu) f_n(y).
\ee
By definition, the Kaluza-Klein modes $\phi^{(n)}$ are mass eigenfunctions of the 4d Laplacian, 
\be
	-\p^2 \phi^{(n)} = m_n^2 \phi^{(n)}. 
\ee	
The wave function profiles $f_n$ are chosen to be orthonormal
\be
	\int_0^{\pi R} dy e^{-2k y} f_n f_m = \delta_{nm}.
\ee
These choices are ensured to be consistent and to completely specify our solutions by standard results in Sturm-Liouville theory, which also guarantees our Kaluza-Klein masses $m_n^2$ are strictly increasing with $n$ to infinity ($m_0^2 < m_1^2 < \cdots$, $\lim_{n\to \infty} m_n^2 = \infty$).  The equation of motion is second order, and the normalization choice of the wave function profiles fixes one of the integration constants. The other is fixed by the arbitrary boundary conditions at each of the branes.

Plugging this into the equations of motion, the wave function profiles $f_n$ satisfy
\be
	\label{scalar:profile}
	-\p_5(e^{-4ky} \p_5 f_n ) + m_\phi^2 e^{-4ky} f_n = m_n^2 e^{-2ky} f_n.
\ee

All that is needed to solve this is the boundary condition for $\phi$, and then we can find the profiles and values of the Kaluza-Klein masses.

\subsubsection{Massless Mode}
\label{scalar:massless}
We will first look for a solution with a mass eigenvalue of $m_0=0$. For the moment setting $b=0$, from (\ref{scalar:profile}) we can find a solution for the wave function profile $f_0$,
\be
	f_0(y) = c_1^{(0)} e^{(2-\sqrt{4+a})ky} + c_2^{(0)} e^{(2+\sqrt{4+a})ky}
\ee
where $c_1$ and $c_2$ are arbitrary constants.

For $a=0$, the solution reduces to
\be
	f_0(y) = c_1^{(0)} + c_2^{(0)} e^{4ky}.
\ee
We can see that to fix Dirichlet or Neumann boundary conditions on the UV and IR branes would force this to be zero. In order to remedy this we must include the boundary mass parameter $b$ so that the boundary conditions are not trivial. The variation of the bulk action requires 
\be
	\delta\phi^{*} \p_5 \phi = 0
\ee
on the boundary. The boundary term we have added adds an additional term proportional to $b$, so the boundary condition either requires that $\phi$ vanishes on both boundaries (and hence everywhere), or that
\be
	\left. f'_0(y) \right|_\text{UV/IR} = b\sigma' \left. f_0(y) \right|_\text{UV/IR}.
\ee

Working this out explicitly, it can be seen that in order to have a solution for $f_0$ other than zero, $b$ must be equal to $2 \pm \sqrt{4+a}$, so we should view the boundary mass as simply a different (but double-valued) parameterization of the bulk mass. It turns out, as we will see, that in the SUSY case this is enforced by the SUSY algebra.  The Breitenlohner-Freedman (BF) bound on the stability of AdS space requires that $a\geq -4$ \cite{Breitenlohner:1982bm}.  This means that the $b$ parameter can have any real value.

With the addition of this boundary mass term, the wave function profile can be solved for explicitly.  The exact solution is stated in (\ref{summary:profile}), but all we need is the overall $y$ dependence, which is
\be
	f_0 =  A e^{bky}
\ee
Putting a profile of this form into the action we get,
\be
	S = -\int d^5x \sqrt{-g} \left[ g^{\mu\nu} \p_\mu \phi \p_\nu \phi^*  + \cdots	 \right]
\ee
\be
	= -\int d^5x \left[ |A|^2 e^{-2\sigma+2bky} \eta_{\mu\nu} \p_\mu\phi^{(0)} \p_\nu {\phi^{(0)}}^* + \cdots \right].
\ee
Because the prefactor is $e^{2ky(b-1)}$, we say that for $b<1$ the zero mode is localized towards the UV brane, for $b>1$, it is localized to the IR brane, and for $b=1$ the zero mode is flat. This corresponds to graphing the wave function profile and taking the warping into account, and seeing which end it's localized to.

\subsubsection{Massive Modes}
Due to the well-ordering of eigenvalues mentioned above, the only other case to look at is when all modes $m_n$, $n>0$, are nonzero. In this case, the general solution for $f_n$ is given in terms of Bessel functions,
\be
	f_n(y) = e^{2ky} \left[ A_n J_{\pm\sqrt{4+a}}\left(e^{ky} \frac{m_n}{k}\right) + B_n Y_{\pm\sqrt{4+a}}\left( e^{ky}\frac{m_n}{k}\right) \right]
\ee
where $A_n$ and $B_n$ are determined by the orthonormality condition of the $f_n$ chosen above, and the boundary conditions.

The Kaluza-Klein masses are determined by imposing Neumann or Dirichlet boundary conditions on the solution and solving for $m_n$ in terms of the zeros of the Bessel functions.  In general, this has no nice analytic solution, but an interesting limiting case is when $R\gg 1/k$, so the extra dimension is large. In this limit the zeros simplify and the masses have the form
\be
	m_n \rightarrow \left( n+\frac{1}{2} \sqrt{4+a} - \frac{3}{4} \right) \pi k e^{-\pi k R}.
\ee
This result is correct for either branch chosen in $b=2\pm\sqrt{4+a}$. Unlike the flat extra dimension case, where the mass scale of the Kaluza-Klein modes is set by $1/R$, it is set by $k e^{-\pi kR}$. The similarity to the IR scale suggests that the KK modes are localized to the IR, which can be seen by explicitly plotting the profiles. Unlike the case of the massless mode, these cannot be arbitrarily localized since they do not each have adjustable parameters to do this \cite{Gherghetta:2010cj}.

\subsection{Fermions}
In 5 dimensions the Dirac algebra is given by\footnote{A general overview of gamma matrices and spinors in $2\leq n\leq 11$ dimensions is given in \cite[Appendix B]{Ortin:2004ms}.}
\be
	\{ \Gamma^M, \Gamma^N \} = 2 g^{MN}
\ee
where $g$ is our curved 5d metric. This can be written in terms of the vielbein $e^M_A$ (defined by $g^{MN} = e^M_A e^N_B \eta^{AB}$) as,
\be
	\Gamma^M = e^M_A \gamma^A
\ee
where the tangent space gamma matrices satisfy the usual relations
\be
	\{ \gamma^A, \gamma^B \} = 2\eta^{AB}.
\ee
For consistency with our conventions before, all of these metrics are `mostly plus.'  These matrices can be chosen to have the representation
\be
	\gamma^\alpha = -i \begin{pmatrix}
		0 & \sigma^\alpha \\
		\overline{\sigma}^\alpha & 0
	\end{pmatrix}, \quad 
	\gamma^5 = \begin{pmatrix}
		1 & 0 \\
		0 & -1 
	\end{pmatrix}
\ee
with $\sigma^\alpha = (1, \sigma^i)$, $\overline{\sigma}^\alpha = (1,-\sigma^i)$, with $\sigma^i$ the Pauli matrices.

In five dimensions, because $\gamma^5$ is part of the Dirac algebra, Lorentz invariant terms cannot depend only on $\gamma^5$, and must involve both left and right handed components. This means that the only allowable spinor representation is in terms of four component Dirac spinors $\Psi$. 

The covariant derivative $D_M = \p_M + \omega_M$ is defined in terms of the spin connection 
\be
	\omega_M = \frac{1}{8} \omega_{MAB} \left[ \gamma^A,\gamma^B \right]
\ee
where $\tensor{\omega}{_M^A_B} = \tensor{e}{^A^N} \tensor{e}{_B_R} \tensor{\Gamma}{_{MN}^R} - \tensor{e}{^A_N} \p_M \tensor{e}{_B^N}$ \cite[ch. 1]{Ortin:2004ms}.  In our metric the spin connection works out to be \cite{Gherghetta:2010cj}
\be
	\omega_M = \left( -\frac{k}{2} e^{-ky}\gamma_\mu \gamma^5, 0 \right).
\ee

The bulk fermion action in 5d is
\be
	\label{fermions:bulkaction}
	S_\Psi = -\int d^5x \sqrt{-g}\left[ \frac{1}{2} \left(\overline{\Psi} \Gamma^M D_M \Psi - D_M \overline{\Psi} \Gamma^M \Psi \right) + m_\Psi \overline{\Psi}\Psi\right].
\ee
Here we have not added a boundary mass. We can see from the $\mathbb{Z}_2$ parity of the field $\Psi$, a boundary Dirac mass term would vanish since $\overline{\Psi}\Psi$ vanishes on the boundaries. A boundary Majorana mass term can be added in the bulk or boundary, but such a term would either be inconsistent or would be equivalent to our choice of even/odd boundary conditions \cite{Contino:2004vy}. A bulk Majorana mass would be inconsistent with supersymmetry so we will not consider it.

We can again define a bulk mass parameter $c$, $m_\Psi = ck$. Even though there are no 5d representations in terms of 5d Weyl spinors, we can decompose the Dirac spinor in terms of two components as if it was composed of 2-component Weyl spinors.  
\be
	\Psi = 
	\begin{pmatrix}
		\psi_+ \\
		\psi_-
	\end{pmatrix}, 
	\Psi_+ =
	\begin{pmatrix}
		\psi_+ \\
		0
	\end{pmatrix},
	\Psi_- =
	\begin{pmatrix}
		0 \\
		\psi_-
	\end{pmatrix}
\ee
so that $\Psi = \Psi_+ + \Psi_-$ and $\Psi_\pm = \pm \gamma_5 \Psi_\pm$. (We label these by $\pm$ instead of L/R for now.)

It is still the case that our representation is not a product of left and right handed components, meaning well-defined chiralities will not exist.  However, later when we impose boundary conditions states of definite chirality will exist. This does not violate Witten's theorem on Dirac operators mentioned in the introduction, because the boundary conditions follow from the existence of orbifold fixed points, which violate the smoothness assumptions of that theorem. 

Putting this into the action and varying it gives us the equations of motion
\be
	\label{fermion:eom}
	e^{ky} \gamma^\mu \p_\mu \Psi_\mp \pm \p_5 \Psi_\pm + (c \mp 2)k \Psi_\pm = 0.
\ee
As before, we can separate variables,
\be
	\Psi_\pm(x^\mu,y) = \sum_{n=0}^\infty \Psi_\pm^{(n)} (x^\mu) f_n^\pm(y)
\ee
and $\Psi^{(n)}_\pm$ are the Kaluza-Klein modes satisfying $\gamma^\mu\p_\mu \Psi_\pm^{(n)} = - m_n \Psi^{(n)}_\mp$. The profiles are easily shown to satisfy
\be
	\pm \p_5 f_n^\pm + (c\mp 2) k f_n^\pm = m_n e^{ky} f_n^\mp.
\ee

\subsubsection{Massless Mode}

The equation of motion for $m_0=0$ is
\be
	\gamma^\mu\p_\mu \Psi^{(0)}_\pm = 0.
\ee
These solutions have definite helicity. This is not concerning, since as we saw before we expect massless modes to follow as a result of particular boundary conditions.

Solving for the profile $f_0$ with $m_0=0$,
\be
	f_0^\pm(y) = d_0^\pm e^{(2\mp c)ky}
\ee 
with one arbitrary constant $d_0^\pm$ this time, since we had a first order equation.  

From our discussion of fields that are even/odd under the $\mathbb{Z}_2$ parity in section \ref{bulkfields:expansions}, we can see that $\Psi_\pm$ have opposite parities. This means that one of the fields is fixed on the boundary, but the equations of motion require such a field to be identically zero.  This means $\Psi$ has right or left handed massless particles, but not both!  Our 4d low-energy theory will come from integrating out massive Kaluza-Klein modes, so this ensures that the low energy theory will be a chiral theory.

This can equivalently be seen by noting that the 5d variational principle requires that the variation on the boundary vanish in addition to the variation in the bulk. The boundary variation can be written, in terms of $\Psi_\pm$,
\be
	0 = \delta S  = \delta S_\text{bulk} + \frac{1}{2} \int d^4x \left[\sqrt{-g} \left( \overline{\Psi}_- \delta\Psi_+ + \delta\overline{\Psi}_+ \Psi_- - \overline{\Psi}_+ \delta\Psi_- - \delta\overline{\Psi}_- \Psi_+   \right) \right]_\text{UV/IR}.
\ee
From this point of view, we must add a boundary action term to the 5d action,
\be
	\label{fermions:boundaryaction}
	S_4 = \frac{1}{2} \int d^4x \sqrt{-g} \left( \overline{\Psi}_- \Psi_+ + \overline{\Psi}_+ \Psi_- \right).
\ee
Then, as we saw from our boundary conditions earlier, for the massless mode, this boundary term is zero. So adding this boundary term is equivalent to imposing our orbifold boundary conditions. This action can also seen to be equivalent to choosing general modes, which have one chirality vanishing on the IR, but not the UV (so that this boundary term only exists on the UV brane), and imposing a separate UV boundary condition. Additional terms could of course be added to this action if they vanish consistently with chosen boundary conditions, but we will only consider the minimal consistent action.

Putting this solution into the action as we did for scalars (taking for definiteness $\Psi_-=0$, making the opposite choice changes $c$ to $-c$)
\be
	S = -\int d^5x \left[ e^{2(c-\frac{1}{2})ky} \overline{\Psi}^{(0)}_+ \gamma^\mu\p_\mu \Psi^{(0)}_+ + \cdots \right] 
\ee
and we see that for $c>1/2$, fermions are localized to the UV, for $c<1/2$ to the IR, and for $c=1/2$ they are flat.

\subsubsection{Massive Modes}

Modes with $m_n\neq 0$ again give us a Sturm-Liouville equation for each chirality as can be seen from (\ref{fermion:eom}) by eliminating one chirality. Of course, massive modes do not have well-defined chiralities here, and each massive mode is a Dirac fermion. Our wave function profiles are again given by Bessel functions
\be
		f_n(y) = e^{\frac{5}{2}ky} \left[ A_n J_{\pm\sqrt{4+a}}\left(e^{ky} \frac{m_n}{k}\right) + B_n Y_{\pm\sqrt{4+a}}\left( e^{ky}\frac{m_n}{k}\right) \right]
\ee
with constants $A_n$, $B_n$ determined from orthonormality and boundary conditions.  Here, the correct choice of normalization is different than the scalar case, and has weight given to us by standard results in Sturm-Liouville theory,
\be
	\int dy e^{-3ky}f^\pm_m f^\pm_n = \delta_{mn}.
\ee

Because the form of $f$ is nearly identical to the scalar case, masses can only differ though the difference in the overall weight and $A_n$ and $B_n$ factors. This means that formally the calculation of masses is identical, and still cannot be done analytically in the general case. In the large $R$ limit we examined before, one can show that in the case of Dirichlet boundary conditions, to allow for a chiral zero mode,
\be
	m^\pm_n \rightarrow \left( n+\frac{1}{2} \left|c\pm\frac{1}{2}\right| - \frac{1}{4} \right) \pi k e^{-\pi k R}.
\ee
So that the only differences here from the scalar case are the bulk mass parameter, the factor of $1/4$ instead of $3/4$, and the $\pm$ coming from using either allowing either $\Psi^+$ or $\Psi^-$ to satisfy Dirichlet boundary conditions.

We note that the massive states, not having definite chirality, are Dirac fermions, so that the fermion spectrum consists of a single massless Weyl fermion and an infinite tower of massive Kaluza-Klein Dirac fermions.

\subsection{Vector Fields}

A bulk vector field $A_M$, which we will assume for simplicity is a U(1) gauge field, has action
\be
	S = -\int d^5x \sqrt{-g} \frac{1}{4} F^{MN}F_{MN}.
\ee	
Gauge fields should always be massless to preserve gauge symmetry, so we do not attempt to add any bulk or boundary masses.  A convenient choice of gauge is $A_5=0$, $\p_\mu A^\mu=0$. With this gauge our equations of motion become
\be
	\eta^{\alpha\beta} \eta^{\mu\gamma} \p_\beta F_{\alpha\gamma} + \eta^{\mu\delta}\p_5 ( e^{-2ky} \p_5 A_\delta) = 0.
\ee

Again, we expand our field,
\be
	A_\mu(x^\nu,y) = \sum_{n=0}^\infty A_\mu^{(n)}(x^\nu) f_n(y).
\ee
This time, the Kaluza-Klein fields satisfy the Proca equation $\eta^{\alpha\beta}\p_\alpha F^{(n)}_{\beta\mu} = m_n^2 A^{(n)}_\mu$. Substituting our expansion into the equations of motion,
\be
	-\p_5(e^{-2ky} \p_5 f_n ) = m_n^2 f_n.
\ee
Once again, this is a Sturm-Liouville equation. This time our profiles have no weight
\be
	\int dy f_n f_m = \delta_{nm}.
\ee

\subsubsection{Massless Mode}

Using the profile's equation with $m_0=0$ we find
\be
	f_0(y) = a_0 + a_1 e^{2ky}
\ee
with arbitrary constants $a_1$ and $a_2$. Imposing Dirichlet boundary conditions requires that $a_1=a_2=0$, so we impose Neumann ones. In this case we simply require that $f'_n=0$ on the boundaries.  So,
\be
	f_0 = \frac{1}{\sqrt{\pi R}}.
\ee
Writing the action in the flat metric,
\be
	S = \int d^5x \frac{1}{4}\frac{1}{\pi R} F^2 + \cdots
\ee	
we can see that this corresponds to having a flat wave function profile.  So we see that we cannot localize massless vector gauge bosons in the bulk. This is not surprising, since we have introduced no bulk or boundary mass which is what localized profiles in the previous cases.

\subsubsection{Massive Modes}

Due to the massive equations having the same Sturm-Liouville form as the past two cases, we know that the general form of the wave function profiles is again
\be
		f_n(y) = e^{ky} \left[ A_n J_{1}\left(e^{ky} \frac{m_n}{k}\right) + B_n Y_{1}\left( e^{ky}\frac{m_n}{k}\right) \right].
\ee

In the large $R$ limit the excited masses can easily be found to be
\be
	m_n \rightarrow \left(n \mp \frac{1}{4}\right) \pi k e^{-\pi k R}
\ee
which are again localized to the IR brane. The $-$ is chosen for Neumann boundary conditions (with a massless zero mode) and the $+$ for Dirichlet (with no massless zero mode).

\subsection{Tensor Fields}

Gravitons are the only tensor field we will consider here, and it can be analyzed by perturbing the metric $g_{MN}$ by a small $h_{MN}$. Because we want to consider particles which correspond only to 4d tensor particles, we will only consider the $\mu\nu$ components of $h$. A complete discussion of the scalar (radial) fluctuations of the metric would be a discussion of the Goldberger-Wise mechanism which was outlined earlier in section \ref{Goldberger-Wise}.  The vector mode is eliminated by our orbifold condition, so we will not have to worry about it.

We will assume the metric has the form
\be
	ds^2 = e^{-2\sigma} ( \eta_{\mu\nu} + h_{\mu\nu} ) dx^\mu dx^\nu + dy^2
\ee
where $|h_{\mu\nu}(x^\alpha,y)|\ll 1$.  As in the usual analysis of gravitational waves we will use the transverse-traceless gauge $\p_\mu h^{\mu\nu}=h^\mu_\mu=0$ (for a review of gravitational waves, see \cite{Ortin:2004ms} or Misner, Thorne, and Wheeler). With this metric our bulk action (see (\ref{bulk:gravity})) becomes
\be
	S = -\int d^5x \frac{M_5^3 e^{-2ky}}{4} \left( \p_\alpha h_{\mu\nu} \p^\alpha h^{\mu\nu} + e^{-2ky} \p_5 h_{\mu\nu} \p_5 h^{\mu\nu} \right)
\ee
which has equations of motion analogous to the usual 4d ones,
\be
	-\p^2 h_{\mu\nu} e^{2ky} + \p_5 (e^{-4ky} \p_5 h_{\mu\nu}) = 0.
\ee
Expanding the metric perturbation,
\be
	h_{\mu\nu}(x^\alpha,y) = \sum_{n=0}^\infty h^{(n)}_{\mu\nu}(x^\alpha) f_n(y).
\ee
As in the previous cases, we have a Sturm-Liouville equation, and here the normalization condition is
\be
	\int dy e^{-2ky} f_n f_m = \delta_{nm}.
\ee

Boundary conditions, just as in the vector case, will either be $f_n(\text{UV/IR}) = 0$ or $f'(\text{UV/IR}) = 0$, because we have no bouk or boundary masses.

\subsubsection{Massless Mode}

This leads to equations for the wave function profiles that look just like the vector field case,
\be
	f_0 = a_1 + a_2 e^{4ky}.
\ee
Dirichlet conditions require $f_0=0$, and Neumann force it to be constant, just as in the vector case. However, when we put this back in the action,
\be
	S \propto \int d^5x e^{-2ky} \p_\alpha h^{(0)}_{\mu\nu} \p^\alpha {h^{(0)}}^{\mu\nu} + \cdots
\ee
so the massless mode is localized to the UV. This means that gravity naturally probes the UV scale, and will be exponentially weaker in the IR. This gives a nice possible explanation of why gravity is so weak compared to the other forces. In the bulk, all couplings, including the gravitational coupling, can be of order one, but the graviton zero mode is always exponentially small in the 4d low energy effective theory!

\subsubsection{Massive Modes}

Unsurprisingly, the solution for $f_n$ is given in terms of Bessel functions.
\be
	f_n(y) = e^{2ky} \left[ A_n J_{2}\left(e^{ky} \frac{m_n}{k}\right) + B_n Y_{2}\left( e^{ky}\frac{m_n}{k}\right) \right].
\ee
In the large $R$ limit masses are seen to be
\be
	m_n \rightarrow \left( n + \frac{1}{2} \mp \frac{1}{4} \right) k\pi e^{-\pi k R}
\ee
where the $-$ sign is taken for Neumann boundary conditions and the $+$ for Dirichlet.  Again, we see that the Kaluza-Klein modes are localized to the IR.

In this case, we can easily find the contribution of these modes to the usual gravitational potential. This calculation was first done for the case of one brane (which is equivalent to our large $R$ case) in \cite{Randall:1999vf} (compare its (12) to our expansion). The potential between two particles located at a separation $r$ on the IR brane is modified by the potentials from the sum of each of the KK modes. Each massive mode contributes a modification due to the Yukawa-like interchange of massive particles,
\be
	\Delta V_n = \frac{G_5 m_1 m_2}{r} \frac{e^{-mr}}{r} \frac{m_n}{k}.
\ee
We can approximate the total effect from all the massive states with an integral,
\be
	\Delta V = \frac{G_4}{k} \frac{m_1 m_2}{r} \int_0^\infty \frac{e^{-m r}}{r}\frac{m}{k} dm = \frac{G_4 m_1 m_2}{k^2 r^3}
\ee
so the Newtonian potential is modified to
\be
	V = \frac{G_4 m_1 m_2}{r}\left( 1 + \frac{1}{r^2k^2} \right).
\ee
For a small $R$, this will have corrections proportional to $1/R$ which encode the finite size of the extra dimensions, and the discreteness of the Kaluza-Klein spectrum.

\subsection{Summary of Fields}
\label{summaryoffields}

\begin{center}
\begin{tabular}{ | l | l | l | l |}
	\hline
	Field									& Zero Mode 									& Bulk Mass	and														&  Massive Mode\\
												&	Profile											&	Zero Mode Localization									&  Localization\\
  \hline
  Scalar $\phi$ 				& $e^{(1\pm\sqrt{4+a})ky}$		& $b>1$ UV, $b<1$ IR, $b=1$ flat					&	IR \\
  Fermion $\Psi$   			& $e^{-(c\pm\frac{1}{2})ky}$	& $c>1/2$ UV, $c<1/2$ IR, $c=1/2$ flat		& IR \\
  Vector $A_\mu$				& $1$													& 0, flat																	& IR \\
  Graviton $h_{\mu\nu}$ & $e^{-ky}$										& 0, UV																		& IR \\
  \hline
\end{tabular}
\end{center}

In each case we have introduced a mass parameter (that may be zero) which determines the localization of the zero mode field. Each field has a tower of massive Kaluza-Klein excitations which are localized to the IR, whose masses are calculated from the boundary condition that allows for a massless zero mode.  For scalars, we have also introduced a boundary mass $b$, which by consistency is required to be related to the bulk mass $a$ through $b=2\pm\sqrt{4+a}$. Fermions have a massless Weyl zero mode, and massive Dirac modes.  Alternatively, we may choose the boundary conditions which do not allow for the zero mode we want and get only massive Dirac particles with no corresponding massless mode.

Scalars have two branches, given by the minus sign chosen in the double-valued parameterization of $b$ in terms of $a$. Fermions also have two solutions, obtained by keeping the left or right handed solution for massless Weyl fermions.

Each field has an expansion in terms of Kaluza-Klein modes and differential equations for wave function profiles given by Sturm-Liouville theory. By inspection, and looking at the definition of the Sturm-Liouville equation we can see that all of the fields and expansions we have written can be summarized in a single set of equations \cite{Gherghetta:2000qt}.

The general equation of motion is given by:
\be
	\left[ e^{2\sigma} \eta^{\mu\nu} \p_\mu\p_\nu + e^{s\sigma} \p_5(e^{-s\sigma}\p_5) -M_\Phi^2 \right] \Phi(x^\mu,y) = 0
\ee
where $\Phi =\left\{ \phi, e^{-2\sigma}\Psi_\pm, A_\mu, h_{\mu\nu} \right\}$, $M^2_\Phi = \left\{ ak^2 + b\sigma'', c(c\pm 1)k^2 \mp \sigma'', 0, 0 \right\}$, $s=\{4, 1, 2, 4\}$.

Wave function profiles are normalized as
\be
	\label{summary:normalization}
	\int dy e^{(2-s)\sigma} f_n(y) f_m(y) = \delta_{nm}
\ee
which satisfy, though the equations of motion,
\be
	\left[ -e^{s\sigma} \p_5 (e^{-s\sigma} \p_5) + \hat{M}^2_\Phi \right] f_n = e^{2\sigma} m_n^2 f_n
\ee
where $\hat{M}$ is $M$ without boundary terms, which are included instead through boundary conditions.

The general solution for $f_n$ is given by
\be
	\label{summary:profile}
	f_n(y) = \frac{e^{s\sigma}}{N_n} \left[ J_\alpha\left(e^\sigma \frac{m_n}{k} \right) + b_\alpha Y_\alpha\left( e^\sigma \frac{m_n}{k}\right) \right]
\ee
with 
\be
	\alpha^2 = \left(\frac{s}{2}\right)^2 + \frac{\hat{M}^2}{k^2}.
\ee
The normalization constant $N_n$ is fixed by the normalization of $f_n$ and the constant $b_\alpha$ is fixed by the boundary conditions.

Neumann boundary conditions (even $\Phi$):
\be
	\left( \frac{df_n}{dy} - r\sigma' f_n \right)_\text{UV/IR} = 0
\ee
where $r=\{b,\mp c, 0, 0\}$.  In this case,
\be
	\label{masses:even}
	b_\alpha(m_n) = - \frac{(r+s/2)J_\alpha\left(\frac{m_n}{k}\right) + \frac{m_n}{k} J'_\alpha\left( \frac{m_n}{k} \right)}{(r+s/2)Y_\alpha\left(\frac{m_n}{k}\right) + \frac{m_n}{k} Y'_\alpha\left( \frac{m_n}{k} \right)}
\ee
\be
	b_\alpha(m_n) = b_\alpha(m_n e^{\pi k R}).
\ee
These two equations together allow us to solve for the masses $m_n$ in terms of Bessel function zeros. In this case, that gives us, in the large $R$ limit,
\be
	m_n \sim \left( n + \frac{|\alpha|}{2} - \frac{3}{4} \right)\pi k e^{-\pi k R}.
\ee

Dirichlet boundary conditions (odd $\Phi$)
\be
	\left. f_n(y)\right|_\text{UV/IR} = 0
\ee
and
\be
	\label{masses:odd}
	b_\alpha(m_n) = -\frac{J_\alpha(m_n/k)}{Y_\alpha(m_n/k)}
\ee
\be
	b_\alpha(m_n) = b_\alpha(m_n e^{\pi k r}).
\ee
Again, we can find the masses,
\be
	m_n \sim \left( n + \frac{|\alpha|}{2} - \frac{1}{4} \right)\pi k e^{-\pi k R}.
\ee

Because both of these scales are set by $ke^{-\pi k R}$, we define the Kaluza-Klein mass scale $M_\text{KK} = k e^{-\pi k R}$. This scale is generically the scale at which the low energy effective theory will start to see Kaluza-Klein modes. By standard results in Sturm-Liouville theory we expect in the general case for our Kaluza-Klein masses $m_n^2$ to be strictly increasing with $n$ to infinity ($m_0^2 < m_1^2 < \cdots$, $\lim_{n\to \infty} m_n^2 = \infty$). This can also be seen from an analysis of the Bessel functions in the general solutions above. In particular, this guarantees only one massless zero mode for any particular field.

\section{The Standard Model in the Bulk}
\label{SMinbulk}

In our model, the 4d Standard Model fields correspond to the low energy effective theory we obtain by integrating out the Kaluza-Klein modes. This means the Standard Model particles should correspond to the light modes of our field expansion. Choosing the field content of the bulk is then a simple matter. We simply upgrade a Standard Model field to a 5d field that has a corresponding zero mode component.

Getting the Standard Model couplings, and describing electroweak symmetry breaking, will take more work.  A lengthy discussion of the $SU(2)_L\times U(1)_Y$ part of the Standard Model in the bulk, assuming the IR scale is the weak scale, can be found in \cite{Casagrande:2008hr,Bauer:2009cf}. Since we will ultimately be interested in building the holographic dual to this theory, which will have a much higher IR scale, we will not study this phenomenology in detail. Instead, we will focus on aspects which do not depend strongly on the specific IR scale.

\subsection{Yukawa Couplings}
\label{yukawa}
Now that we know what all of our bulk fields look like, we can write the Standard Model content in the bulk. To discuss Yukawa couplings, we need to upgrade each Standard Model Weyl fermion $\psi_L$ to a bulk Dirac fermion $\Psi_L$ (the $L$ is now just a label on the Dirac fermion that tells us which zero mode to choose), and the Standard Model Higgs to a bulk scalar $H$.  As we saw in section section \ref{scales}, introducing a scalar field confined to the IR brane introduced a warp factor $e^{-\pi k R}$ in the action. If we introduce additional fields which are not localized to the IR, then they will have smaller warp factors associated with them, so we expect interaction terms in the 5d action like
\be
	\overline{\Psi}_L H \Psi_R
\ee 
will, when written as canonically normalized terms, have Yukawa couplings proportional to exponential factors.  Thus we should be able to generate hierarchical Yukawa couplings by placing fields at appropriate positions in the bulk.

Confining the Higgs to the IR for the moment, our bulk action now contains a term
\be
	\int d^5x \sqrt{-g} \tilde{\lambda}_{ij} \left[ \overline{\Psi}^L_i \Psi^R_j + \text{h.c.}\right] H \delta(y-\pi R)
\ee
\be
	= \int d^4 x \lambda_{ij} \left[ \left(\overline{\Psi}^L_i\right)^{(0)}_+ \left(\Psi^R_j\right)^{(0)}_- H + \text{h.c.} + \text{KK modes} \right].
\ee
Where $\lambda$ defines the 4d Yukawa coupling, and $\tilde{\lambda}$ is the 5d coupling. We know the zero mode profile is proportional to $e^{-(c_i-2)ky}$ for each fermion $\Psi^{L,R}_i$. Using (\ref{summary:normalization}) and (\ref{summary:profile}) we can find the overall wave function profile normalization,
\be
	f^{(0)}_i (y) = \sqrt{\frac{(1-2c_i)k}{e^{(1-2c_i)\pi k R}-1}} e^{(2-c_i)ky}
\ee
Where the $i$ index comes with an $L+$ and the $j$ with a $R-$. Putting this into our action and equating terms, we can solve for the 4d couplings,
\be
	\lambda_{ij} = \tilde{\lambda}_{ij}k \frac{1}{N_{i} N_{j}} e^{(1-c_i-c_j)\pi k R}
\ee
where the normalizations $N_{i,j}$ come from the prefactor of $f^{(0)}$.

We can see that we can get an exponential hierarchy of couplings by varying the $c$s. To get an estimate of how large these couplings are, let's assume $\tilde{\lambda}k \sim 1$. Let's look at a couple of limits.

If $c_i=c_j=c$ then,
\be
	\lambda_{ij} = \frac{(1/2 - c) e^{k \pi R}}{e^{k\pi R} - e^{2 c k \pi R}}.
\ee
To reproduce the electron coupling of $10^{-6}$ in this case, we'd need $c\approx 0.69$ for $kr \approx 10$. So the zero component of the electron field (the massless field before getting a mass from the Higgs) is UV-localized. This is what we expect, since a UV localized fermion has a small overlap with the IR localized Higgs, and so the exponential factor between them in the action should have a large negative exponent. We can see by putting in a larger coupling, we would have smaller values of $c$. For example, for a coupling of order one, we'd need $c=-0.5$, which is strongly IR localized.

Similarly, we can place the Higgs on the UV brane by changing the delta function to $\delta(y)$, and find couplings in the case $c_i=c_j=c$,
\be
	\lambda_{ii} = \frac{1-2c}{e^{\pi k R (1-2c)} - 1}
\ee
which is exponentially small in the IR ($c<1/2$) and large in the UV ($c>1/2$). For the flat case of $c=1/2$ the coupling is $\lambda = 1/(\pi k R)$. Technically, we should be worried about this, because in the absence of SUSY the Higgs receives large radiative corrections if it lives on the UV brane. However, soon we will be adding SUSY and can place the Higgs anywhere we like.

These can be combined into one formula by noting that the overal normalizations $N_i$ do not depend on the localization of the Higgs, and this only comes into the couplings from the exponentials that show up in the action. This means we can write a general $\lambda$ for a Higgs localized at $y^* = 0$ or $\pi R$ as
\be
	\label{y5d:general}
	\lambda_{ij} = \frac{k \tilde{\lambda}_{ij} }{N_i N_j} e^{(1-c_i-c_j)ky^*}.
\ee

\subsection{Gauge Couplings}
\label{gaugecouplings}

Standard Model fermions enjoy gauge coupling universality.  That is, all fermion flavors have the same couplings to Standard Model gauge bosons. One might worry that the same effect that causes an exponential Yukawa hierarchy will cause couplings to be non-universal. So let's look at a 5d U(1) gauge field $A_\mu$ coupled to a 5d fermion $\Psi_i$. This action contains
\be
	\int d^5x \sqrt{-g} g_5 \left[\, \overline{\Psi}_i(x,y) \Gamma^\mu A_\mu(x,y) \Psi_i(x,y) \right] 
\ee
\be
	= \int d^4x g_4 \left[ \overline{\Psi}^{(n)}_i(x) \gamma^\mu A^{(0)}(x) \Psi_i^{(m)}(x) \cdots \right].
\ee
Using the profiles we calculated before, and doing the $y$ integral using the orthonormality conditions, we can see that the first line reduces to
\be
	g_5\int dy e^{-3ky} f_0^A f_n^{\Psi_i} f_n^{\Psi_j} = g_5 \frac{\delta_{mn}}{\sqrt{\pi R}}
\ee
so our 4d flavor couplings are universal and are scaled by a factor of $1/\sqrt{\pi R}$ compared to the 5d couplings. This applies to all of the Standard Model zero modes.  We should not be surprised at this result, because 4d gauge invariance is preserved, so we would expect the 4d low-energy EFT to have universal couplings. (A similar argument can be constructed to show that gravity couples universally as well.)

However, because the Kaluza-Klein modes of the gauge bosons are not flat, but are localized towards the IR, they will introduce small non-universal couplings. We can find these by replacing the zero mode of $A^\mu$ in the expression above with a massive mode, and looking at the coupling to the fermion zero modes,
\be
	g_5 \int dy e^{-3ky} f_n^A f_0^{\Psi_i} f_0^{\Psi_j} 
\ee
however this is generally a complicated integral of Bessel functions that has no nice solutions. They have been numerically calculated in \cite{Gherghetta:2010cj, Casagrande:2008hr, Bauer:2009cf} where it is shown that for $c>1/2$ the couplings become very close to universal, and for $c<1/2$ they can differ substantially. This is not entirely unexpected since the massive gauge boson states are localized to the IR.

\subsection{Conclusions}

At this point, we have developed enough tools that we can take any conceivable Standard Model or beyond the Standard Model interaction, upgrade it to 5d, put mode expansions into the action, and calculate couplings, masses, and suppression factors for higher-dimension operators, such as those responsible for FCNCs, and come up with constraints and experimental signs of extra dimensions. Many papers have done this, and a few places to start looking at this are \cite{Csaki:2004ay, Gherghetta:2010cj, Casagrande:2008hr, Bauer:2009cf, Kim:2003pc, ArkaniHamed:1999dc}. Typically these are all calculated in terms of wave function overlap integrals such as the ones we have just found. These are all done by assuming the UV scale is of order the Planck scale, and the size of the extra dimensions is such that the IR scale is the TeV scale. 

What we will be more interested in doing from this point, however, is introducing SUSY, and using the basic tools we have developed to understand a different realization of this model. One where we can use AdS/CFT to help us understand the 4d theory, and where the IR scale is not necessarily the TeV scale. 

\section{Supersymmetry}

Now we can introduce supersymmetry and see how it effects this setup. One might worry that introducing branes to the bulk explicitly breaks SUSY by breaking translation invariance, but it has been shown that spaces with domain walls can still be supersymmetric by properly accounting for boundary terms \cite{Bergshoeff:2000ii}. Because we are in AdS space instead of flat space, however, there is one important difference between ordinary SUSY models and our model. The momentum operator $P$ does not commute with SUSY charges and $P^2$ is not a Casimir operator. This means that fields of different mass can belong to the same supermultiplet.

In five dimensions, there are eight supercharges, so our bulk fields are $\mathcal{N}=2$ supermultiplets. We will show that the orbifold condition breaks half of this supersymmetry. So our model can accommodate the MSSM fields as components of bulk $\mathcal{N}=1$ representations.

\subsection{Gravity Supermultiplet}

The on-shell gravity multiplet contains the vielbein $e^A_M$, the graviphoton $B_M$ (which our orbifold conditions will require to be zero), and two symplectic-Majorana\footnote{A symplectic-Majorana spinor is defined by taking a set of Dirac spinors $\Psi^i$, $i=1,\ldots 2n$, and imposing a reality condition on the whole set: $\overline{\Psi}^i=\Psi_i^c = \Omega_{ij}\psi^{jc}$, where $\Omega^2 = - $Id, and Id is the identity operator.} gravitinos $\Psi^i_M$, where the $i$ labels the SU(2) index of the SUSY algebra automorphism group. The action looks like the usual 5d SUGRA action plus a boundary cosmological constant term for each brane. The form of the action is not important to us, so we will not write it explicitly. In this case, as in \cite{Bergshoeff:2000ii} the bulk and boundary cosmological constants are related by the domain wall plus bulk SUSY algebra. This means that the condition $\Lambda_\text{bulk} = -k \Lambda_\text{IR} = k\Lambda_\text{UV}$ which before was an additional condition required by consistency, is now enforced by SUSY.

The orbifold parity requires that our symplectic-Majorana spinors satisfy the following relation under the $\mathbb{Z}_2$ symmetry:
\be
	\eta^i(-y) = (\sigma_3)^{ij} \gamma_5 \eta^j(y)
\ee
where $\eta$ is the infinitesimal generator of SUSY transformations.

The condition that SUSY is unbroken is $\delta \Psi^i_M = 0$, which translates to the Killing equation
\be
	D_M \eta^i = -\frac{\sigma'}{2} \gamma_M (\sigma_3)^{ij} \eta^j.
\ee
The orbifold condition enforces that in addition to this
\be
	\gamma_5 \eta^i = (\sigma_3)^{ij}\eta^j.
\ee
This condition relates some half of the symmetry transformations to other half (the relations above can be seen as breaking the SUSY generators into `even' and `odd' ones), and so breaks half of the $\mathcal{N}=2$ supersymmetry.  So the orbifold theory only realizes $\mathcal{N}=1$ 4d SUSY.

\subsection{Vector Supermultiplet}

There is a vector supermultiplet $\mathcal{V}$ consisting of a gauge field $V_M$, a symplectic-Majorana spinor $\lambda^i$, and a real scalar $\Sigma$ in the adjoint representation of the gauge group associated with $V$.

If we consider a U(1) gauge group for simplicity, the action is
\begin{align}
	S_\mathcal{V} = -\frac{1}{2} \int d^5x \sqrt{-g}[ &\frac{1}{2g^2} F^2 + \\	
	& \p_M\Sigma\p^M\Sigma + m^2_\Sigma \Sigma^2 + \nonumber\\
	& i\overline{\lambda}^i\Gamma^M D_M\lambda^i + i m_\lambda \overline{\lambda}^i (\sigma_3)^{ij}\lambda^j] \nonumber.
\end{align}
In flat space the masses $m_\Sigma$ and $m_\lambda$ would have to be zero, since the gauge boson's mass is zero. However, as we have mentioned, in AdS space that is not the case and the masses may be different.

The SUSY transformations for a vector supermultiplet are
\begin{align}
	\delta V_M &= -i \overline{\eta}^i \gamma_M \lambda_i \nonumber\\
	\delta \Sigma &= \overline{\eta}^i \lambda^i \\
	\delta \lambda^i &= (-\Sigma^{MN}F_{MN} + i \gamma^M \p_M \Sigma)\eta^i - 2 i \sigma' \Sigma \sigma_3^{ij} \eta^j
\end{align}
where $\eta$ is the Killing spinor defined above.

These relations enforce that
\be
	m^2_\Sigma = - 4k^2 + 2 \sigma''
\ee
\be
	m_\lambda = \frac{1}{2}\sigma'.
\ee
Recall from before, that to ensure a consistent solution for the nonvanishing scalar zero mode, the first condition was required. Here it is a result of the SUSY algebra.

We can find the mass parameters $a$, $b$, and $c$ for these scalar and spinor fields just as was done in the nonsupersymmetric case by expanding in modes. In this case we have
\be
	a=-4, \quad b = 2\pm\sqrt{4+a} = 2
\ee
and
\be
	c = \frac{1}{2}
\ee
along with the vector mass parameter, which is zero.

There is an interesting relation that results from this. Recall that the way mass parameters show up in Kaluza-Klein masses is through $\alpha$,
\be
	\alpha^2 = \left(\frac{s}{2}\right)^2 + \hat{M}^2.
\ee
For the vector, scalar, and fermion we find, 
\be
	\alpha_V^2 = 1 + 0 = 1, \quad	\alpha^2_\Sigma = 4-a = 0, \quad \alpha^2_\lambda = \frac{1}{4} + c (c\pm 1) = 1,0.
\ee
This tells us how to choose boundary conditions: we can consistently choose $(V_\mu,\lambda^1)$ to be even and $(\Sigma, \lambda^2)$ to be odd.  If this is done, the first pair has a massless mode which forms a 4d $N=1$ multiplet, and the second pair will not have a massless mode. Additionally, if we look at the even/odd mass formulas from (\ref{masses:even}, \ref{masses:odd}), and using $\alpha_{V,\lambda^1}/2 - 3/4 = \alpha_{\Sigma,\lambda^2} - 1/4$ so in the limit where the radius is large, all Kaluza-Klein masses in this multiplet will be the same.  Investigating the full even/odd mass formulas, we can show that this equality of Kaluza-Klein masses holds in general, not just for the large $R$ limit.

\subsection{Hypermultiplet}
\label{susy:hypermultiplet}
This theory has a hypermultiplet $\mathcal{H}$, which contains two complex scalar fields $H^i$ and a Dirac fermion $\Psi$. This has an action
\be
	S_\mathcal{H} = -\int d^4x \sqrt{-g} \left( \left|\p_M H^i\right|^2 + i \overline{\Psi}\Gamma^M D_M \Psi + m^2_{H_i}\left|H^i\right|^2 + im_\Psi \overline{\Psi}\Psi\right).
\ee
The SUSY transforms in this case are 
\be
	\delta H^i = \sqrt{2}i\epsilon^{ij} \overline{\eta}^j \Psi
\ee
\be
	\delta\Psi = \sqrt{2}\left[ \Gamma^M \p_M  H^i \epsilon^{ij} - \frac{3}{2}\sigma' H^i (\epsilon\sigma_3)^{ij} - m_\Psi \overline{\Psi}\Psi \right]\eta^j.
\ee
As in the vector multiplet case, this determines the masses of each of these fields.
\be
	m^2_{H^{1,2}} = \left( c^2 \pm c - \frac{15}{4} \right) k^2 + \left( \frac{3}{2} \mp c \right) \sigma''
\ee
\be
	m_\Psi = c \sigma'.
\ee
From this we can see that the bulk mass parameter of the fermion is $c$, and the mass parameters of the scalar are
\be
	a= c^2 \pm c - \frac{15}{4}
\ee
and
\be
	b= \frac{3}{2} \mp c.
\ee
As before we can calculate the $\alpha$s, 
\be
	\alpha^2_{H^1,\Psi_+} = \left(c+ \frac{1}{2} \right)^2, \quad \alpha_{H^2,\Psi_-} = \left( c - \frac{1}{2}\right).
\ee
We see that SUSY has related not only the scalar bulk and boundary mass parameters to each other, but to the fermion mass parameter as well. Putting these results into the even/odd mass formulas (\ref{masses:even}, \ref{masses:odd}) and simplifying the Bessel functions, we can see that the Kaluza-Klein masses are equal for all particles in the multiplet.

The boundary conditions we saw before for each of these fields require that only the even fields here have a massless mode. So, like for the vector multiplet, the massless sector, containing $H^1$ and $\Psi_L$ form an $\mathcal{N}=1$ massless chiral multiplet. If odd boundary conditions are chosen, we choose the other sign in $\alpha$ and there is no massless mode.

\subsection{Summary}

In terms of the analysis done in the non supersymmetric case, adding supersymmetry has done two things: it determines the bulk and boundary masses of each particle in a multiplet in terms of each other particle in the multiplet, and it ensures that all Kaluza-Klein masses of all particles in the multiplet are the same. If superpartners are visible in a collider, the results of section \ref{SMinbulk} will produce distinctive predictions for the couplings of superpartners.  If $M_\text{KK}$ is small enough to be accessible by a collider, this also produces a distinctive mass spectrum for the Kaluza-Klein states each supermultiplet.

Adding supersymmetry also has the same stabilizing effects in 5d as in the 4d MSSM (which from the point of view of the bulk theory are inherited from the 5d theory). 

\section{The 5d MSSM}
\label{5dmssm}

Now we can place all MSSM supermultiplets in the bulk in the same way we could place all Standard Model fields in the bulk. Because with SUSY we do not have to worry about radiative corrections destabilizing the Higgs mass if it is not on the IR brane, we can place it anywhere in the bulk, but for now we will confine it to the IR or the UV, although it is easy to construct other models where this is not the case.  

To look at what can be done, we will construct two different models, with particle content localized in different places. In both of these cases, the IR scale is chosen to be the TeV scale, and the UV scale is chosen to be the Planck scale. These choices aren't necessary, but allow for a straightforward application of what we have constructed to phenomenologically reasonable models. Soon, we will abandon the idea that the IR scale is related to the weak scale, and investigate the relationship to AdS/CFT. 

\subsection{Partly Supersymmetric Standard Model}
Although the MSSM solves the hierarchy problem, there are potentially large FCNC and CP violating terms from the soft mass Lagrangian. A simple solution to this problem is to have all MSSM scalar masses at the Planck scale, which can be done while still solving the hierarchy problem in a warped model by breaking SUSY on UV brane, which is placed at the Planck scale.

This model is constructed by breaking SUSY on the UV brane through a spurion field, and keeping the bulk and IR supersymmetric. The IR brane is chosen to be at the TeV scale, and the UV brane is chosen to be at the Planck scale. Matter is localized according to the table below. This model was first proposed by Gherghetta and Pomarol in \cite{Gherghetta:2003wm}. We will see that this model contains only some supersymmetric particles, so we refer to it as the Partly Supersymmetric Standard Model (PSSM).

\begin{center}
\begin{tabular}{ | l | l | l |}
	\hline
	UV 		& Bulk		& IR \\
	\hline
	Broken SUSY 	& SM gauge bosons & Higgs \\
	    					& SM fermions    & \\
  \hline
\end{tabular}
\end{center}

Supersymmetry breaking can be parameterized by a spurion field living on the UV brane, $\eta = \theta^2 F$ with $F\sim M_P^2$, along with the a term leading to gaugino masses
\be
	\int d^2\theta \frac{\eta}{M_P^2} \frac{1}{g_5^2} W^\alpha W_\alpha \delta(y) + \text{h.c.}.
\ee
This gives a gaugino mass $m_\lambda\sim F/M_P \sim M_P$. The gravitino will also receive a Planck scale mass. So both the gaugino and gravitino will decouple from the low energy theory. SUSY can also be broken for squarks and sleptons with a UV term,
\be
	\int d^4\theta \frac{\eta^\dagger\eta}{M_P^4} k  S^\dagger S \delta(y)
\ee
which leads to soft scalar masses which are also of order the Planck scale, decoupling squarks and sleptons from the low energy theory as well. All potentially light superpartner zero modes (except for the Higgs) in the 5d theory are given a Planck scale mass in this way, reducing the low energy theory to the usual SM fields (aside from the Higgs, which will be discussed in a moment). 

Kaluza-Klein modes of 5d fields, as we have seen, are always IR localized, giving them redshifted masses compared to their UV-localized zero modes. This results in a distinctive spectrum with nonsupersymmetric zero modes, and approximately supersymmetric KK modes which have TeV scale masses (recall that the KK mass scale is $M_\text{KK} = ke^{-\pi k R} \approx$ TeV, in this case).

Because the Higgs is on the IR brane, it will not see a SUSY breaking UV term, but will only see SUSY breaking terms from one loop interactions with gauge fields. A soft one-loop Higgs mass from this effect will be of order
\be
	m_H \sim \sqrt{\frac{\alpha}{4\pi}} M_\text{KK}.
\ee
The KK modes play an important role in keeping the Higgs mass natural. The difference between KK boson and fermion modes ends up canceling quadratic divergences that would drive the Higgs mass up. This is what must happen because in the 5d theory the excited KK modes are contained in 5d multiplets. But from the low energy theory's point of view, this seems like an unexpected cancellation. So the KK tower is what keeps the Higgs mass natural even though SUSY is broken at the Planck scale.

We have not completely specified the Higgs sector, however. As in the MSSM, the higgsino generates gauge anomalies that we have to cancel. This can be done in a number of different ways, such as having two Higgs doublets whose anomalies cancel each other, having one Higgs doublet and moving a Standard Model fermion to the IR to cancel the anomaly, or constructing a model where the higgsino is a Standard Model particle localized on the IR brane. A detailed analysis of each of these cases can be found in \cite{Gherghetta:2003wm}.

\subsection{Warped MSSM}
In this setup the warp will be used to generate the TeV scale in the IR, which can naturally generate TeV scale particle masses, and is the SUSY breaking scale. We localize all fields except gauge fields on the UV brane. Again, we choose the UV brane to be the Planck scale and the IR brane to be the TeV scale. This model was investigated in detail in \cite{Gherghetta:2000kr}.

\begin{center}
\begin{tabular}{ | l | l | l |}
	\hline
	UV 		& Bulk		& IR \\
	\hline
	Higgs 				& SM gauge bosons		& Broken SUSY \\
	SM Fermions		& and superpartners				 			& \\
								&  & \\
  \hline
\end{tabular}
\end{center}

Confining all matter to the UV naturally suppresses dangerous higher dimension operators by powers of $M_P$. 

Supersymmetry is broken explicitly through choosing `twisted' boundary conditions which are inconsistent with supersymmetry on the IR brane, but allowing the bulk and UV to be supersymmetric.  Before we defined fermions to satisfy
\be
	\psi(-y) = \pm \gamma_5 \psi(y) 
\ee
with the same sign at each boundary. Recall, this breaks half of the supersymmetry. Alternatively, we can define the signs to be different at each boundary 
\be
	\psi(0) = \gamma_5 \psi(0), \quad \psi(\pi R) = - \gamma_5 \psi(\pi R)
\ee
which prevents a Killing spinor from being defined, and thus breaks SUSY. In flat space this boundary condition corresponds to the Scherk-Schwarz mechanism \cite{Gherghetta:2000kr}, but in this warped model it does not because there is no smooth limit in which supersymmetry is restored.  Interestingly, in the flat space model these boundary conditions are equivalent to having a spurion superfield with a SUSY breaking $F$-term, but in this theory this is not the case.  Here, such a term would be suppressed by the warp factor and leads to a similar, but different spectrum than the one considered here.  In a similar model constructed in \cite{Barbieri:2000vh} it is shown that these boundary conditions correspond to a $S^1/(\mathbb{Z}_2\times\mathbb{Z}_2)$ orbifold.

This boundary condition corresponds to choosing an $\alpha = |c+1/2|$ fermion, which has no massless mode, which can be seen from the general solution for the mass spectrum given above. Using the general mass formula we have derived, this means the gaugino zero mode receives a mass 
\be
	\label{gauginomass}
	m_\lambda \sim M_\text{KK}/\sqrt{kR}
\ee 
because it is IR localized, and the UV localized gravitino receives a mass 
\be
m_{3/2} \sim M_\text{KK}^2/k
\ee
Using $M_\text{KK} \sim $ TeV, we find $m_\lambda\sim .1 $ TeV, and $m_{3/2} \sim 10^{-3}$ eV.

Gauge interactions will communicate SUSY breaking from the IR brane to the UV brane through one loop effects, which should give scalar masses at one loop of order $m_\text{scalar}^2 \sim \alpha/(4\pi) M_\text{KK}^2$. This is the same order as the Higgs mass in the last section even though SUSY was broken on the opposite brane.

\subsection{Conclusions}

We have seen in these two models that soft masses of the reasonable sizes can be produced. Importantly, we see that the Higgs does not have to live on a brane associated with the weak scale, it can even be all the way up at the Planck scale and masses of the right size can still be generated. As we saw in section \ref{SMinbulk} the Higgs can correctly produce an exponential hierarchy and flavor universal couplings even if placed in the UV. This will be important to us in the next section when the IR scale may be higher than the weak scale. 

Many papers in the literature can be found that continue in this direction using these and similar models to calculate corrections to Standard Model physics, through various electroweak precision tests, $S$ and $T$ parameters, couplings, and other effects.  Instead of investigating the detailed consequences of these models, we will now move on to looking at 4d holographic duals of our 5d theories and see if this sheds any additional light on the problem of constructing phenomenologically reasonable 4d models.

\section{AdS/CFT}

A remarkable and unexpected result in string theory is that ten dimensional type IIB string theory on backgrounds of the form $\text{AdS}_d \times M_{10-d}$ has a dual description in terms of an ordinary ($d-1$)-dimensional superconformal gauge theory, which is interpreted as living on the `boundary' of the AdS space. The duality states that these two theories are equivalent to each other, and states, observables, correlation functions, and dynamics, can be calculated in one theory and transformed to corresponding results in the other theory.

There are many excellent reviews of AdS/CFT \cite{D'Hoker:2002aw,Petersen:1999zh,Klebanov:2000me}. What we will be interested in is outlining the basics of this correspondence that will be useful to us later, and developing a modified version of this correspondence that works in a slice of AdS, instead of the whole space.

\subsection{Overview}

In the case where $M_{10-d}=S^5$, this correspondence is between $\text{AdS}_5 \times S^5$ and $\mathcal{N}=4$ $SU(N)$ superconformal Yang-Mills on $\mathbb{R}^{3,1}$.  We can quickly check that the symmetries of each of these theories are identical. 

The $\mathcal{N}=4$ supersymmetry on the CFT side has an associated $SU(4)$ $R$-symmetry, which corresponds to the isometry group of the sphere $S^5$, which is $SO(6) \cong SU(4)$. On the AdS side, there are $N$ units of 5-form flux running through the five sphere, corresponding to the rank of the gauge group on the CFT side. It can also be seen that the conformal symmetry of the CFT, $SO(3,1)$ is the group of isometries of AdS. 

If $g_\text{YM}$ is the CFT coupling, $k$ is the AdS curvature radius, and $l_s$ is the string length, then the AdS/CFT correspondence states that these parameters are related by
\be
	\frac{1}{k^4 l_s^4} = 4\pi g^2_\text{YM} N.
\ee
We can immediately see that the limit where string theory is approximated by classical gravity, $l_s \ll 1/k$, corresponds to $4\pi g^2_\text{YM} N \gg 1$, which is the condition that the CFT is strongly coupled.  Additionally, one requires that in the string theory, nonperturbative states remain heavy, which means that the string coupling $g_s$ must be small, implying that the duality holds in the large $N$ limit. The correspondence can then be seen as an example of a duality between theories with strong and weak coupling, analogous to Seiberg duality between strongly and weakly coupled $SU(N)$ gauge theories.

The gravity description is given in terms of the $d$-dimensional fields living in AdS. The CFT is described in terms of some set of operators. This means there must be some correspondence between bulk fields $\Phi(x^M)$ and CFT operators $\mathcal{O}$. Because the CFT lives on the boundary of the bulk, the boundary value of fields,
\be
	\left. \Phi(x^M) \right|_\text{AdS boundary} = \phi_0(x^\mu)
\ee
act as sources for the CFT operators $\mathcal{O}$. 

\subsection{AdS/CFT in a Slice of AdS}
In the usual AdS/CFT correspondence, there is only one boundary, the boundary of AdS space. It is simple to modify the discussion above to have fields which are specified instead of on the AdS boundary, on some other boundary inside of AdS (the UV), and impose boundary conditions on second boundary in AdS (the IR).

To write down the dual theory on the new boundary, first we want to integrate over bulk fields $\Phi$, constrained to have a particular UV boundary value $\phi_0=\Phi(x^\mu,y_\text{UV})$, and define this as the effective action \cite{Contino:2004vy},
\be
	\mathcal{Z}[\phi_0] = \int_{\Phi=\phi_0} \mathcal{D}\Phi e^{i S[\Phi]} \equiv e^{iS_\text{eff}[\phi_0]}.
\ee
This can be done explicitly by solving the 5d equations of motion, and substituting the result back into the action.  Additionally, IR boundary conditions must be chosen consistently with the 5d variational principle, as discussed in section \ref{bulkfields:expansions}.

Next, we integrate over all possible UV boundary values $\phi_0$, including the any UV boundary action pieces,
\be
	\label{AdSCFT:Z5d}
	\mathcal{Z}_\text{5d} = \int \mathcal{D}\phi_0 e^{i(S_\text{eff}[\phi_0]+S_\text{UV})}.
\ee

The AdS/CFT correspondence states that this is equivalent to calculating, in a strongly coupled large $N$ SCFT with gauge group $SU(N)$,
\be
	\mathcal{Z}_\text{CFT} = \int \mathcal{D}\Phi_\text{SCFT} e^{i(S_\text{SCFT}+\phi_0 \mathcal{O})}.
\ee

This tells us CFT that correlation functions can be calculated directly in terms of the 5d effective action,
\be
	\label{AdSCFT:correlation}
	\left< \mathcal{O}\mathcal{O}\cdots\mathcal{O}\right> = \frac{\delta^n \ln\mathcal{Z}_\text{5d} }{\delta\phi_0\cdots\delta\phi_0}
\ee
which allows us to calculate all $n$-point functions of one theory in terms of the other.

To be careful, this correspondence should be discussed in terms of the regularized / renormalized AdS/CFT correspondence, which describes the correspondence in terms of a regularized / renormalized CFT coupled to a local theory for the boundary fields, containing UV and IR terms. For our purposes, we will not worry about any renormalization or regularization, and implicitly assume this has been taken care of for the results we state. For a careful treatment of scalar fields in the regularized AdS/CFT correspondence, see \cite{PerezVictoria:2001pa}.

Because the isometries of AdS correspond to the conformal group in the field theory, the $y$-coordinate, which tells us our `depth' along the AdS throat corresponds to an energy scale in the CFT. Translation invariance in the bulk then corresponds to renormalization group flow in the field theory. If we add domain walls, this `ends' the space, and introduces UV or IR cutoffs in the field theory, breaking conformal invariance.

\subsubsection{UV Cutoff}

A UV cutoff in the CFT is a scale $\Lambda_\text{UV}$, which corresponds to a position $y=y_\text{UV}$ in the bulk, that explicitly breaks conformal invariance. Adding only a UV brane keeps the AdS space infinite in the IR direction, so moving to the IR quickly restores approximate conformal symmetry. This means that the conformal symmetry breaking at the UV scale must be by irrelevant operators.  

Introducing the UV cutoff causes the source field $\phi_0$ to become a dynamical field. We can see this because the AdS theory, the field's value on the boundary is dynamical. This requires the introduction of the term $S_\text{UV}$ in (\ref{AdSCFT:Z5d}) above.

\subsubsection{IR Cutoff}

Introducing a cutoff at $y=y_\text{IR}$ corresponding to a CFT scale $\Lambda_\text{IR}$ also breaks conformal symmetry, by cutting off the remaining half of AdS.  In the CFT, introducing explicit breaking in the IR would correspond to the addition of a relevant deformation in the CFT, but this operator would be associated through the AdS/CFT correspondence to a scalar with negative mass. But no such particle is needed for consistency in the bulk, so the breaking must be spontaneous \cite{Rattazzi:2000hs}.  

This breaking is not entirely spontaneous, however. This can be seen by recalling that the location of the IR brane is determined by the vev of the radion field.  This radion is massless and its dual is interpreted as the Goldstone boson associated with the spontaneous breaking of dilations in the CFT, the dilaton. Stabilizing the radion introduces a mass, and explicitly breaks conformal symmetry in the IR. This means the radion/dilaton is now a pseudo-Goldstone boson, analogous to how spontaneous chiral symmetry breaking in QCD results in massless pions, which are Goldstone bosons, and are given a mass by a small explicit breaking.

A nontrivial consequence of this is that the presence of the IR cutoff will, though the introduction of the scale $\Lambda_\text{IR}$, introduce a mass gap in the theory, and the CFT will have a discrete spectrum with masses of order $m\sim kn$ for $n=1,2,\ldots$.

\subsubsection{Embedding into String Theory}

Although there are no known embeddings of RSI into string theory, there are very similar solutions in spaces like $AdS_5 \times T^{1,1}$, such as Klebanov-Strassler \cite{Klebanov:2000hb}, where fluxes close off the IR end of the geometry, and the UV end of the geometry is a Calabi-Yau. If the RSI model is embeddable in string theory, it would be expected to have a similar solution, and all of the results quoted above could be made more formal, but such a procedure is not required to justify our results.

\subsection{Holography for Scalar Fields}

We would now like to use this setup to calculate the 4d dual theory in the case of a single bulk scalar field \cite{Gherghetta:2010cj}.  As outlined above, we can proceed by putting the solution to the equations of motion into the action.  The solution we have found in terms of modes easily motivates the functional form of the general solution, so we can start by looking at the solution in terms of it.   From (\ref{summary:profile}) we know that the bulk field has an expansion
\be
	\label{holo:scalar}
	\phi(x^\mu,y) = \sum_{n=0}^\infty \phi_n(x^\mu) f_n(y),
\ee
which we can put into the scalar action and use to find the effective action $S_\text{eff}$.  Using the separation of variables solution in modes, and Foruier transforming in four dimensions, we can see that transforming the whole function $\phi$ without expanding in modes should have the form
\be
	\phi(p,y) = \phi(p) g(y) h(\tilde{p})
\ee
where $\Phi(p)$, $g$, and $h$ are determined by the equations of motion for a scalar field, and $\tilde{p} = p e^{-ky}$ is the redshifted momentum. Putting this solution back into the action, we can find that the effective action has the form
\be
	S_\text{eff} = \frac{k}{2}\int \frac{d^4p}{(2\pi)^4} \phi(p)\phi(-p) F(\tilde{p}_\text{UV},\tilde{p}_\text{IR}).
\ee
Where $\tilde{p}_\text{UV/IR} = \left. p e^{-ky} \right|_\text{UV/IR}$ are the redshifted momenta on the boundaries. We note that the redshift factor at the UV brane is of course one, but we keep the label `UV' to keep track of where terms originated, which is important to keep track of boundary contributions. Using the boundary conditions that $\phi$ satisfies determines the exact form of $F$. If we use the boundary conditions chosen in section \ref{scalar:massless}, $F$ has the form
\be
	F(\tilde{p}_\text{UV},\tilde{p}_\text{IR}) = \mp i \tilde{p}_\text{UV} f(\nu\mp 1) f(\nu)
\ee
\be
	f(m)=J_{m}(i\tilde{p}_\text{UV}) - Y_{m}(i \tilde{p}_\text{UV}) \frac{J_\nu(i \tilde{p}_\text{IR})}{Y_\nu(i\tilde{p}_\text{IR})}
\ee
where $\nu = \alpha\pm 1$, which represents the two different signs we signs we had in our solution for bulk scalar masses (recall, $b=2\pm\sqrt{4+a}$).

The self-energy can be calculated explicitly from (\ref{AdSCFT:correlation}) using the $F$ determined above. 
\be
	\Sigma(p) = \int d^4x e^{-ip\cdot x} \frac{\delta^2 S_\text{eff}}{ \delta( \phi(x,0))\delta(\phi(0,0))}
\ee

The self energy $\Sigma(p)$ is identical to the CFT correlation function $\left<\mathcal{O}\mathcal{O}\right>$ up to analytic terms, and the poles are the masses of our Kaluza-Klein states. The analytic terms in (the properly regulated / renormalized) $\Sigma(p)$, unlike in the usual AdS/CFT case, as a result of the UV cutoff, correspond to kinetic terms for our CFT source field $\phi_0 = \phi(x,0)$. This means that a modified mass spectrum must be obtained by using the action
\be
	S = S_\text{UV} + S_\text{eff}
\ee
as in (\ref{AdSCFT:Z5d}).  This gives us a new mass spectrum which contains our physical states. This spectrum can be found to have no massless poles in the case where $\nu_- = -(1-\alpha)$ and a massless pole in the case where the other branch $\nu_+ = 1+\alpha$ is chosen.

Examining the two point function $\left<\mathcal{O}\mathcal{O}\right>$ with counterterms we find \cite{PerezVictoria:2001pa} that the scaling dimension is 
\be
	\label{scalar:dim}
	\text{dim} \mathcal{O} = 2+\sqrt{4+a} = 2 + |b-2|,
\ee
which is true for either sign of $\nu$ chosen above. 

For the $\nu_-$ branch, with no massless poles, below the cutoff scale $\Lambda\sim k$ the dual 4d Lagrangian has the form
\be
	\label{scalar:L}
	\mathcal{L} = -Z_0 (\p_\mu \phi_0)^2 + \frac{\omega}{\Lambda^{\nu_-}} \phi_0 \mathcal{O} + \mathcal{L}_\text{CFT}
\ee
where $Z_0$ and $\omega$ are running couplings which can be determined through matching this Lagrangian to the self energy $\Sigma(p)$. This Lagrangian shows mixing through $\phi_0\mathcal{O}$ term of the massless source field $\phi_0$ and the CFT bound states. There are three cases to look at to understand this mixing:\\
{\bfseries Irrelevant Coupling}. This corresponds to $\nu_->0$, which we can see from section \ref{summaryoffields} corresponds to a UV-localized mode. For irrelevant coupling, the mixing will not be substantial, which is what is expected for a mode localized to the UV. This means that the bulk zero mode field is dual to the CFT source field, in the limit that mixing is ignored completely.\\
{\bfseries Relevant Coupling}. This corresponds to $\nu_-<0$, which is an IR localized mode. Here the mixing is large, and here the dual of the bulk zero mode is a mixture the of source field $\phi_0$ and of CFT bound states.\\
{\bfseries Marginal Coupling}. This corresponds to $\nu_-=0$, which is a flat mode in the bulk. In this case the mixing is not as large as for a relevant operator, but is still not negligible, and the bulk zero mode is a mixture of the source field and CFT bound states.

The $\nu_+$ branch contains a massless mode, which corresponds to a massless scalar mode in the CFT. The renormalized self energy contains a constant term, which corresponds to a mass term in the dual Lagrangian, which has the form
\be
	\mathcal{L} = - Z_0 (\p_\mu \phi)^2 + m_0^2\phi^0 + \frac{\omega}{\Lambda^{\nu_+-2}}\phi_0 \mathcal{O} + \mathcal{L}_\text{CFT}
\ee
where the running couplings can again be obtained through matching. The mass term here ends up being of order $k$.

A similar analysis can be carried as above and as before, relevant and marginal operators are IR localized and composite, and irrelevant ones are UV localized and elementary.

\subsubsection{Renormalization Group}
\label{ren:scalar}

We can investigate the coupling $\omega$ in (\ref{scalar:L}) by rescaling the coupling to a dimensionless coupling
\be
	\xi(\mu) = \frac{\omega}{\sqrt{Z(\mu)}} \left( \frac{\mu}{\Lambda} \right)^\gamma
\ee
with anomalous dimension $\gamma$.  The RG equation for this CFT is
\be
	\mu \frac{d\xi}{d\mu} = \gamma\xi + c \frac{N}{16\pi^2} \xi^3.
\ee
Here, $c$ is an order 1 constant that comes from the CFT's contribution to the wavefunction renormalization.  The first term comes from the source/CFT mixing term $\phi\mathcal{O}$.  Relating the normalization of $\omega$ to the Lagrangians above, we see we should identify $\gamma$ with $\pm\nu_\pm = \alpha\pm 1$.  The RG equation is easy to solve, and we can see that for $\pm\nu_\pm=\gamma<0$, which corresponds to a relevant coupling and IR localized fields gives us,
\be
	\xi \sim \frac{4\pi}{\sqrt{N}} \sqrt{|\gamma|}
\ee
and for $\gamma>0$, which corresponds to irrelevant operators,
\be
	\xi \sim \frac{4\pi}{\sqrt{N}} \sqrt{\gamma}\left( \frac{\mu}{\Lambda} \right)^\gamma
\ee
This is exactly what we expect! Relevant operators (corresponding to IR localized fields) have a coupling that is non-negligible and constant at low energies, and irrelevant ones (UV localized fields) are small except at high energies.

\subsection{Holography for Fermions}

This can be repeated without too much trouble for fermions.  Recall that the bulk action for fermions (\ref{fermions:bulkaction}) has a corresponding boundary term (\ref{fermions:boundaryaction}) required by the variational principle in the bulk. In our case, because we would like to consider source fields as UV boundary values, let us consider a general fermion (not necessarily one corresponding to a massless zero mode) where as discussed earlier, the IR boundary conditions fix one chirality to be zero but does not fix the UV boundary condition. This means the boundary term (\ref{fermions:boundaryaction}) can be written only on the UV brane,
\be
	\label{holofermions:boundaryaction}
	S_{\text{UV}} = - \frac{1}{3} \int d^4x \left[\sqrt{-g} \left( \overline{\Psi}_L \Psi_R + \overline{\Psi}_R \Psi_L \right) \right]_\text{UV}.
\ee

Using this, along with the bulk action, we can integrate out the bulk degrees of freedom. As before, we do this by solving the equations of motion, and substituting them into the action. This is made easy in this case because we have imposed boundary conditions through the boundary action, and the on-shell bulk action vanishes, so we only need to deal with (\ref{holofermions:boundaryaction}). 

Inspired by the decomposition used in the scalar case, we look for a solution that has the form
\be
	\Psi_{L,R}(p,y) = f(p,y,y_\text{UV},y_\text{IR}) \Psi(p)_{L,R}.
\ee
We can choose this function $f$ to have the form
\be
	f(p,y,y_\text{UV},y_\text{IR}) = \frac{f_{L,R}(p,y)}{f_{L,R}(p,y_\text{UV})}.
\ee
To simplify our calculation, we can change coordinates to use the metric \cite{Contino:2004vy} 
\be
	ds^2 = a(z) \left( \eta_{\mu\nu} dx^\mu dx^\nu - dz^2 \right), \quad z\in[z_\text{UV},z_\text{IR}].
\ee
The $f_{L,R}$ are required to satisfy the equation
\be
	\left( \p_z + 2 \frac{\p_z a(z)}{a(z)} \pm a(z) m_\Psi \right) f_{L,R} = \pm p f_{R,L}.
\ee
In terms of these functions, the Dirac equation requires
\be
	\slashed p \Psi_R(p) = p \frac{f_R(p,z_\text{UV})}{f_L(p,z_\text{UV})} \Psi_L(p).
\ee
Putting this solution into the action and rescaling $\Psi_L(p) \rightarrow \Psi_L(p)/a(z_\text{UV})^{3/2}$ to get canonical kinetic terms, we get,
\be
	S_\text{boundary} = \int \frac{d^4p}{(2\pi)^4} \overline{\Psi}_L(p) \Sigma(p) \Psi_L(p)
\ee
where
\be
	\Sigma(p) = a(z_\text{UV}) \frac{p}{\slashed p} \frac{f_R(p,z_\text{UV})}{f_L(p,z_\text{UV})}.
\ee
As before, this self-energy corresponds to the CFT correlation function $\left<\mathcal{O}\mathcal{O}\right>$ up to analytic terms.

Using the definition of $f_{L,R}$, their exact expressions can be found in terms of Bessel functions \cite{Contino:2004vy}. At this point, the analysis is the same as for the scalar case, and we find that the scaling dimension of $\mathcal{O}$ is
\be
	\text{dim} \mathcal{O} = \frac{3}{2} + \left| c \pm \frac{1}{2} \right|.
\ee
This is in contrast to \cite{Contino:2004vy} which has this solution, but only for the $+$ branch. 

Now we can write Lagrangians for the dual theory and look at relevant and irrelevant operators for different values of $c$.  There are three cases to look at $c\geq 1/2$, $-1/2\leq c \leq 1/2$, $c\leq -1/2$.  We will examine only the plus branch case, the minus case is qualitatively similar.

\subsubsection{{ \slshape c} \textgreater { 1/2}}%$c\geq 1/2$}
\label{c>1/2}
In this case the 4d theory contains a dynamical source field $\Psi_L(p)$ with Lagrangian below the scale $\Lambda\sim k$ \cite{Contino:2004vy} 
\be
	\mathcal{L} = \mathcal{L}_\text{CFT} + Z_0 \overline{\Psi}_L i \slashed{\p} \Psi_L + \frac{\omega}{\Lambda^{\alpha-1}}(\overline{\Psi}_L \mathcal{O} + \text{h.c.}) + \xi \frac{\overline{\mathcal{O}}\slashed{p}\mathcal{O}}{\Lambda^{2\alpha}} + \cdots
\ee
where $\mathcal{L}_\text{UV}$ comes from any additional terms in the UV Lagrangian, and $Z_0$, $\omega$, and $\xi$ are running couplings, and the $\cdots$ are other deformations to the CFT suppressed by powers of $\Lambda$. From before, $\alpha=c+1/2$.  In this case, $c>1/2$ corresponds to an irrelevant coupling and $c=1/2$ to a marginal one. 

We can see what happens when we take the limit that the cutoff goes to infinity. If we rescale the field $\Psi_L \rightarrow \Lambda^{\alpha-1} \Psi_L$, then our Lagrangian contains terms
\be
	\mathcal{L} = \mathcal{L}_\text{CFT+UV} + \frac{Z_0}{\Lambda^{2(\alpha-1)}} \overline{\Psi}_L i \slashed{\p} \Psi_L + \omega (\overline{\Psi}_L \mathcal{O} + \text{h.c.}) + \cdots.
\ee
Simultaneously taking the large $\Lambda$ and small IR cutoff limits, the kinetic term and the $\cdots$ terms vanish, the UV Lagrangian vanishes, and we are left with
\be
	\mathcal{L} = \mathcal{L}_\text{CFT} + \omega(\overline{\Psi}_L \mathcal{O} + \text{h.c.}).
\ee
This is the Lagrangian of a fixed external source field probing a CFT. This is what we expect to happen from the bulk interpretation, as this limit removes the branes which recovers the usual AdS/CFT correspondence.

A careful analysis of UV cutoff effects and CFT masses can be found in \cite{Contino:2004vy}.

\subsubsection{-1/2 \textless { \slshape c} \textless { 1/2}}%$-1/2 \leq c \leq 1/2$}
\label{-1/2<c<1/2}
In this case, dim $\mathcal{O} \leq 5/2$, so the coupling is now either relevant or marginal.  This means that the coupling of the field $\Psi_L(p)$ is important at low energies, and will mix with CFT states. All of the other conclusions of the previous section still apply.

\subsubsection{{ \slshape c} \textless { -1/2}}%$c\leq -1/2$}

This case has an interesting difference from the previous two. If we expand our self energy (using the exact form of the $f_{L,R}$ found in \cite{Contino:2004vy}) we find that in the limit the UV cutoff goes to infinity, and the IR cutoff goes to zero,
\be
	\Sigma(p) \sim \slashed{p} \left( \frac{a_{-2}}{p^2} + a_0 + a_2 p^2 + \cdots \right)
\ee
where the $a_n$ are constants which are not important to us, we see that there is a pole in the first term. Because in the limit we have taken the theory is conformal, there is no counterterm to cancel this. This term must correspond to an additional massless field $\chi$ external to the CFT.

This term is seen as analogous to violating the BF bound for scalars, and seems to signal an inconsistency of the theory \cite{Cubrovic:2009ye}, but it can be reinterpreted in terms of a $c>-1/2$ state, and it can be given an interpretation in terms of a Fermi liquid \cite{Cubrovic:2009ye, Cubrovic:2010bf, Iqbal:2011in}. 

\subsubsection{Renormalization Group}
\label{ren:fermions}

Following \cite{Contino:2004vy} let's consider the example of a Yukawa coupling of a right handed quark $q$ and a scalar $\psi$ living on the IR brane coupled to a bulk fermion $\Psi$. We write the interaction term just as we did in section \ref{yukawa},
\be
	\mathcal{L} = \int d^5x \delta(y-\pi R) \sqrt{-g}(\lambda_5 \overline{q}\phi \Psi_L + \text{h.c} ).
\ee
If $c>-1/2$ so that the dual theory is the one described in sections \ref{c>1/2} and \ref{-1/2<c<1/2}, then the 4d running coupling $\lambda(\mu)$ at scale $\mu$ is given by
\be
	\lambda(\mu) = \left(\frac{\mu}{\Lambda}\right)^{\alpha-1} \frac{\omega(\mu)}{\sqrt{Z_0(\mu)}}.
\ee
The RG equation is then
\be
	\label{RGE}
	\mu \frac{d\lambda}{d\mu} = \left(\text{dim}(\mathcal{O}) - \frac{5}{2}\right)\lambda + c \frac{N}{16\pi^2} \lambda^3.
\ee
Just as in the scalar case, $c$ is an order 1 constant that comes from the CFT's contribution to the wavefunction renormalization. Using that 
\be
	\gamma=\text{dim}(\mathcal{O}) - \frac{5}{2}=\alpha-1
\ee
we find
\be
	\lambda(\mu) \sim \frac{4\pi}{N} \left(\frac{\mu}{\Lambda}\right)^\gamma  \left[ \frac{\gamma}{1-\left(\frac{\mu}{\Lambda}\right)^{2\gamma}} \right]^{1/2}
\ee
This tells us that the coupling is determined by the anomalous dimension $\gamma$.

This result, that coupling strengths depend on $\gamma$ come form of the RG equation (\ref{RGE}), and does not depend on the specific details of our theory, just that there is a coupling of the form $\lambda\overline{\psi}\mathcal{O}$ between the CFT operator $\mathcal{O}$ and the source field $\psi$.

We can see that for UV localized fields ($\alpha>1$), which we expect to correspond to irrelevant operators and have,
\be
	\lambda(\mu) \sim \frac{4\pi}{\sqrt{N}} \sqrt{\gamma} \left(\frac{\mu}{\Lambda}\right)^\gamma,
\ee
and for IR localized ($0<\alpha<1$) relevant operators, 
\be
	\lambda(\mu) \sim \frac{4\pi}{\sqrt{N}} \sqrt{\gamma}.
\ee
Exactly as in the scalar field case, relevant operators have a coupling that is non-negligible and constant at low energies, and irrelevant ones are small except at high energies.

Considering Yukawa couplings, however, is more interesting. We can see explicitly, that we expect anomalous dimensions to determine Yukawa couplings in the 4d dual theory (and in the 4d effective theory obtained from integrating out CFT modes), and bulk localization to determine the Yukawa couplings in 5d, which determines exponentially suppressed couplings in the 4d effective theory obtained from integrating out massive Kaluza-Klein modes. 

\subsection{Symmetries}
\label{gaugesymmetries}
We briefly note an important consequence of requiring bulk fields to satisfy some gauge symmetry in the bulk. Because all gauge symmetries, being the result of 5d gauge bosons, have a dual CFT interpretation, any bulk gauge symmetry shows up as a global symmetry in the CFT. This can be understood by noting that the CFT without a source contains no massless fields.  Adding a source field weakly gauges this symmetry.

\section{Holographic Mode Expansion}

In the bulk the convenient basis to use is the Kaluza-Klein (mass eigenstate) basis, but the convenient basis to use in the 4d CFT dual is in terms of CFT bound states, not Kaluza-Klein states. The obvious thing to do then, is to expand our fields in a basis corresponding to CFT bound states, match this to the Kaluza-Klein basis, and use it to determine mixings explicitly in terms of these components.  This basis is called the `holographic basis' \cite{Batell:2007jv}.

To do this, in the bulk we will expand a field in terms of the CFT source field and a sum of CFT bound states,
\be
	\Phi(x,y) = \phi^s(x) g_s(y) + \sum_{n=1}^\infty \phi^{(n)}g_n(y).
\ee
As we saw when we introduced the AdS/CFT correspondence, the pure CFT spectrum can be obtained from using Dirichlet boundary conditions in the UV and Neumann (with boundary masses included) in the IR. This means that the source profile $g_s$ is given by our general expression for wave function profiles (\ref{summary:profile}) subject to these boundary conditions.  Additionally, we demand that the wave function profiles are normalized such that they give canonical kinetic terms. Because the form of this is the same as the Kaluza-Klein mode expansion, the form of the solution will still be (\ref{summary:profile}), but the specific terms in the expansion will generally be different because we have imposed different boundary conditions in order to ensure one field corresponds to the CFT source field.

\subsection{Scalar Fields}

For real scalar fields, with bulk mass $b$, the boundary conditions are
\be
	g^{(n)}(\text{UV}) = 0, \quad (\p_5-bk) g^{(n)}(\text{IR}) = 0.
\ee
This means the source field is
\be
	g_s(y) = \left\{ \begin{matrix}
	\sqrt{\frac{2(b-1)k}{e^{2(b-1)\pi k R}-1}} e^{bky}, \quad b<2\\
	\\
	\sqrt{\frac{-2(b-3)k}{e^{-2(b-3)\pi k R}-1}} e^{(4-b)ky}, \quad b>2
	\end{matrix}\right.
\ee
We can see by looking at the exponentials that UV/IR localization corresponds to irrelevant/relevant operators through the operator dimension we found earlier (\ref{scalar:dim}), $\Delta = 2 + |b-2|$.

Putting this expansion into the scalar action with boundary term (\ref{scalar:action}), (\ref{scalar:boundaryaction}), we get
\begin{align}
	S = \int d^5x & \left( -\frac{1}{2}e^{-2ky}g_s^2 (\p_\mu\phi_s)^2 - \frac{1}{2} e^{-4ky}(\p_5 g_s)^2\phi^2_s - \frac{1}{2}ak^2 e^{-4ky}g_s^2 \right. \nonumber\\
	& \left. - bke^{-4ky}g_s^2\phi_s^2(\delta(y)-\delta(y-\pi R)) \right)
\end{align}
where the delta functions came from the boundary action.  We can do the $y$-integrals and get,
\be
	S = S(\phi_s) + S(\phi^{(n)}) + S_\text{mixing}
\ee
\be
	S(\phi_s) = -\frac{1}{2} \int d^4x \left( (\p_\mu \phi_s)^2 - \frac{1}{2}M_s \phi^2_s \right)
\ee
\be
	S(\phi^{(n)}) = -\frac{1}{2} \int d^4x \sum_{n=1}^\infty \left( (\p_\mu \phi^{(n)})^2 - M_n^2 (\phi^{(n)})^2 \right)
\ee
\be
	S_\text{mixing} = \int d^4x \sum_{n=1}^\infty \left( -z_n \p_\mu \phi_s \p^\mu \phi^{(n)} - \mu^2_n \phi_s \phi^{(n)} \right)
\ee
where we have defined constants
\be
	M_s^2 = \frac{e^{2(2-b)\pi k R}-1}{e^{2(3-b)\pi k R}-1} 4(b-2)(b-3)k^2
\ee
\be
	z_n = \int dy e^{-2ky} g_s \phi^{(n)}
\ee
\be
	\mu^2_n = \int dy \left(\p_5 g_s \p_5 g^{(n)} + g_s g^{(n)} m_\phi \right)
\ee
where $m_\phi = ak^2 + b\sigma''$. We have also used that $M_n$ are the CFT mass eigenvalues, defined by the $\phi^{(n)}$.  These latter two integrals, defining $z_n$ and $\mu^2_n$ are wavefunction overlap integrals, analogous to those calculated in sections \ref{yukawa} and \ref{gaugecouplings}.

The mixing term $S_\text{mixing}$ tells us that, as long as $z_n\neq 0$, the basis we have chosen is not orthogonal (although it is normalized properly, since the kinetic terms are canonical). This system can be normalized by writing it in matrix notation and changing the basis \cite{Batell:2007jv}.

\subsection{Matching to the Kaluza-Klein Basis}

To achieve our ultimate goal of understanding CFT mixing in terms of easily understood Kaluza-Klein fields, we will match these two expansions,
\be
	\sum_{n=0}^\infty \phi^{(n)}(x) f_n(y) = \phi_s(x)g_s(y) + \sum_{n=1}^\infty \phi^{(n)} g_n(y).
\ee
We can multiply this by $f_m$ and use its orthonormality condition to collapse the sum on the left hand side,
\be
	\phi^{(n)}(x) = v^n_s \phi_s(x) + \sum_{m=1}^\infty v^{nm}\phi^{(m)}(x)
\ee
with
\be
	v^n_s = \int dy e^{-2ky} f_n(y)(y) g_s(y)
\ee
\be
	v^{nm} = \int dy e^{-2ky} f_n g_m(y).
\ee
Thus we have reduced understanding CFT compositeness to calculating wave function overlap integrals of functions $f$ and $g$ which satisfy specified boundary conditions, and whose form is given by (\ref{summary:profile}).  This is done in detail for bosonic fields in \cite{Batell:2007jv} and for fermions in \cite{Batell:2007ez}.

\subsection{Integrating Out}

Using this correspondence we can see explicitly that we have three slightly different 4d theories we have been looking at.
\begin{itemize}
	\item The 4d effective theory on the IR brane obtained from integrating out massive Kaluza-Klein states.
	\item The 4d effective theory obtained from integrating out massive CFT states.
	\item The 4d effective theory which is dual to the 5d bulk theory.
\end{itemize}
We know that the third 4d theory, the dual, is an exact description of the dynamics in the bulk. Because we can relate KK modes directly to CFT modes, we can see that the first two theories are actually equivalent (although if we have no mass gap in the CFT we may be integrating out massless modes, which is valid only if these modes are weakly coupled to the fields which are not integrated out). 

This means that the procedure outlined in section \ref{SMinbulk} for upgrading a 4d Standard Model field to the zero mode of a 5d bulk field, calculating wave function overlaps, and integrating out KK modes, is exactly the same procedure as writing down the 4d dual theory and integrating out the CFT!  

In section \ref{ann} we will take a 4d extension of the Standard Model that was constructed independently of AdS/CFT, determine its 5d dual, and discuss how the 4d features correspond to the bulk features.

\section{4d Dual of 5d MSSM}

Using our results above, we can translate the models of section \ref{5dmssm} into their duals.

\subsection{Dual of the Partly Supersymmetric Standard Model}

Recall, in this model, SUSY is broken on the UV brane (at the Planck scale), and it is approximately supersymmetric where conformal symmetry is broken on the IR brane (at the TeV scale). This breaks part of the supersymmetry, and we end up with only superpartners to the Higgs at the massless level. The particle content is then: Standard Model particles localized in the bulk (to the UV or IR), a SUSY breaking suprion field in the UV, and a Higgs and Higgsino in the IR.

By our dictionary, the zero modes of massless fields which are UV localized will correspond to CFT source fields. So any 5d Standard Model particles localized to the UV are elementary particles in the 4d theory.  From this point of view, the spurion field can be thought of in the 4d theory as describing SUSY as an accidental symmetry which only appears at low energies.

The zero modes of massless fields localized to the IR are composite, so this theory has a Higgs and Higgsino which are composites of CFT bound states, and are approximately supersymmetric. Kaluza-Klein modes, always localized to the IR, are also CFT composites.

Superpartners of Standard Model fermions, which appear in the spectrum at high energies, will receive Planck scale masses and are UV localized (if their corresponding partners were), so will be elementary.

Gauge bosons are delocalized in the bulk, and will be partly elementary and partly composite, other than the graviton, which is always UV localized, so is predominantly elementary.

Because Standard Model fermions, aside from the Higgs and Higgsino, are UV localized, we can think of this theory as being the usual Standard Model (with gravity), plus a supersymmetric composite Higgs/Higgsino, plus a strongly coupled CFT.

\subsection{Dual of the Warped Supersymmetric Standard Model}

Here, supersymmetry was broken through boundary conditions in the IR, and conformal symmetry is broken in the IR. Both of these can be seen as spontaneous breaking. The Higgs and Standard Model fermions and their superpartners were confined to the UV brane, and so are elementary. Gauge bosons are the same as in the previous case. 

As we saw, gauginos and the gravitino will receive soft masses at tree level, and squarks and sleptons will receive masses at one loop from gauge bosons, which communicate SUSY breaking from the IR to the UV. This is similar to gauge mediation, except with no hidden sector, and where the messengers are charged under the Standard Model gauge group.

In this case, because we have localized fields on the UV brane, and allow only the `messenger' gauge fields to live in the bulk, we can easily understand the 4d interpretation of this `mediation.'  The gaugino has a Dirac mass (because that is the only mass consistent with our twisted boundary conditions) so it mixes with a fermion bound state to become massive,
\be
	\omega \lambda \mathcal{O}_\psi.
\ee
Its mass has to be proportional \cite{Gherghetta:2000kr, ArkaniHamed:2000ds} to
\be
	\sqrt{\frac{g^2 b}{8\pi^2}}
\ee
where $g^2=g_5^2/(\pi R)$ is the 4d coupling and $b=8\pi^2/(g_5^2k)$ is the CFT beta function. Using the relations to the 5d theory, this tells us the mass is proportional to
\be
	\frac{1}{\sqrt{\pi k R}},
\ee
which is what was claimed before. 

We can also see that because the gaugino has a flat profile, $c_\lambda = 1/2$, so that $\text{dim}(\mathcal{O}_\psi) = 5/2$ ($\gamma=1$), and the RG equation has a solution
\be
	\xi^2(\mu) \sim \frac{16\pi^2}{N \ln(k/\mu)} 
\ee
so at the IR scale $M_\text{KK}$, 
\be
	\xi^2(M_\text{KK}) \sim \frac{16\pi^2}{\pi k R}.
\ee
which is exactly the prefactor that appears in (\ref{gauginomass}).

\subsection{The Single Sector MSSM}

We can look at one more 5d model that has some distinctive features in the 4d dual theory \cite{Gabella:2007cp}. Instead of the models above that localize some matter to the IR and UV branes, let us place all the Standard Model content (except for the Higgs) in the bulk, but with some of the fields localized to the IR and some to the UV. As we have seen, Yukawa couplings are generated by wave function overlaps, so this will be able to generate a large hierarchy between the couplings of the first and third generations by placing one at the UV end and one at the IR end, with the third generation closest to the Higgs (so that it will have a large overlap and thus an order one coupling). Let's confine the Higgs to the UV brane so that in the dual theory it is an elementary field. 

SUSY will be broken in the IR, but we will not need to associate this scale with the TeV scale. The IR scale in this model can be thought of as the scale SUSY breaks at, and the scale that conformal symmetry breaks at, and this does not need to be at any particular value. The UV scale can still be thought of as the Planck scale.

\begin{center}
\begin{tabular}{ | l | l | l | l | l |}
	\hline
	UV brane			& 	UV localized		& 	Delocalized		&		IR localized		& IR brane \\
	\hline
	Higgs 				& Third generation	& 	Gauge Bosons	&	First Generation 	&	Broken SUSY \\
  \hline
\end{tabular}
\end{center}

From section \ref{susy:hypermultiplet} we know that a chiral multiplet, like the ones Standard Model fields will be contained in, relates the localization of a fermion with mass parameter $c$ to a scalar with mass parameters
\be
	a = c^2 \pm c - \frac{15}{4}, \quad b=\frac{3}{2}\mp c.
\ee
Similarly, we say that Standard Model gauge bosons, being delocalized, determine gauginos to have $c_\lambda=1/2$ (which is delocalized as well). So localizing the Standard Model zero modes as above determines the localizations, and thus couplings, of the superpartners.  

Because SUSY is broken in the IR, superpartners localized to the IR will see the largest soft masses. In the Standard Model, Yukawa couplings are related to fermion masses and the Higgs vev $v$, so UV localized particles have larger couplings and larger masses, with superpartners that are UV localized and have {\textit small} masses because they are far from the source of SUSY breaking. IR localized particles have small masses and superpartners with large masses, because they are close to the source of SUSY breaking.

We can estimate that the light UV localized superpartners have masses 
\be
	\tilde{m} \sim e^{(\frac{1}{2}-c)\pi k R} M_\text{KK}
\ee
while the heavy IR localized ones have masses 
\be
	\tilde{m} \sim M_\text{KK}.
\ee

This model has an interesting interpretation in terms of its 4d dual. As we've seen, IR localized particles correspond to CFT composites, so the Higgs and third generations are elementary, the second generation is composite, and the first generation is more strongly composite. This is similar to the models of \cite{ArkaniHamed:1997fq, Luty:1998vr} where MSSM states arise as composites of a strongly coupled theory with superpotential terms $\sim Q\overline{U}$, which see SUSY breaking through $F$-terms of $\overline{U}$. These fields $\overline{U}$ act as mediators of the SUSY breaking and communicate SUSY breaking to all of the composite MSSM states, giving their superpartners large soft masses.

Models like this are called `single-sector' because the field $\overline{U}$ is a part of the strongly coupled sector which forms composites that are Standard Model particles, instead of having a distinct `hidden' sector that communicates SUSY breaking \cite{Gabella:2007cp}.

Single-sector-like models can also be constructed directly in $AdS_5 \times M^5$ where the IR brane is obtained through fluxes closing off the end of AdS space, as in \cite{Benini:2009ff}.

\subsection{Summary}

There are a few important lessons to get out of the models we have shown.  First, the IR scale does not need to be the TeV scale, it may be much higher.  The IR scale always corresponds to the conformal symmetry breaking scale in the dual CFT, and in 5d is the scale where Kaluza-Klein modes become important. The EFT below the Kaluza-Klein scale in the 5d theory is equivalent to the 4d EFT obtained from integrating out CFT states in the 4d dual.

We can summarize the scales in the most general case in a diagram.

\begin{center}
\begin{tikzpicture}
\begin{scope}[>=latex]
	\draw (0,11) node {RSI};
	\draw [->, thick] (0,0) -- (0,10);
	\draw (.3,10) node {$E$};
	\draw (-2,9) node {No EFT};
	\draw (-1.5,8.5) node {String scale};
	\draw (-2,8) -- (1,8);
	\draw (1.5,8) node {$\Lambda_\text{UV}$};
	\draw (-2,7) node {5d EFT};
	\draw (-2,6) -- (1,6);
	\draw (2.2,6) node {$\Lambda_\text{IR} \sim M_\text{KK}$};
	\draw (-2,4.5) node {4d EFT};
	\draw [dashed] (-1,4) -- (1,4);
	\draw (2,4) node {TeV scale};
	\draw (-2,2) -- (1,2);
	\draw (1.5,2) node {0};
	
	\draw [dotted] (3.5,0) -- (3.5,10);
	
	\draw (7.2,11) node {Dual theory};
	\draw [->, thick] (7,0) -- (7,10);
	\draw (7.3,10) node {$E$};
	\draw (4.5,9) node {No EFT};
	\draw (5,8.5) node {String scale};
	\draw (5,8) -- (8,8);
	\draw (8.5,8) node {$\Lambda_\text{UV}$};
	\draw (5,7) node {4d CFT};
	\draw (5,6) -- (8,6);
	\draw (9.2,6) node {$\Lambda_\text{IR} \sim M_\text{CFT}$};
	\draw (5,4.5) node {4d EFT};
	\draw [dashed] (6,4) -- (8,4);
	\draw (9,4) node {TeV scale};
	\draw (5,2) -- (8,2);
	\draw (8.5,2) node {0};
\end{scope}
\end{tikzpicture}
\end{center}

The second important lesson is that anomalous dimensions in the 4d CFT correspond to bulk mass parameters in the 5d theory.
\be
	\mu \frac{d\lambda}{d\mu} = \gamma\lambda + c \frac{N}{16\pi^2}\lambda^3
\ee
where $\gamma$ is given by the field's $\alpha$ parameter. This means localization, which in the 5d theory is a trivial observation, is equivalent to something which may be non-trivial in the strongly coupled CFT.

The third thing to remember is that we can write the CFT basis in terms of the KK basis, and use this to understand mixings,
\be
	\phi^{(n)}_\text{5d}(x) = v^n_s \phi_s(x) + \sum_{m=1}^\infty v^{nm} \phi^{(m)}.
\ee

\section{Suppressing Flavor Anarchy}
\label{ann}
Now we can use all of the tools we have developed to analyze the 5d dual of a model which was proposed entirely in terms of 4d language to describe Yukawa hierarchies and CKM mixing angles \cite{Nelson:2000sn}.  This model shows how the observed patterns of couplings and CMK mixings could result from renormalization group flow, where some quarks and leptons acquire large anomalous dimensions though coupling the Standard Model to a CFT, without imposing any flavor symmetries, or other additional symmetries.

First, we will overview the basic features of this model and look at some general predictions, then we will look at a specific $SU(5)$ model, and finally discuss the 5d dual.  We will try to keep results as general as possible.

\subsection{Basic Setup}

Suppose we have a strongly coupled $\mathcal{N}=1$ supersymmetric gauge theory with gauge group $G$, which contains charged matter $Q$ and neutral matter $X$, and has a superpotential $W(Q,X)$. Assume this theory becomes conformal in the IR.

Unitarity and gauge invariance requires that all gauge invariant operators $\mathcal{O}$ of this theory have dim $\mathcal{O}\ge 1$. This means any such operator has a positive anomalous dimension, e.g., for the operator $X$, $\gamma_X = 2(\text{dim}(X)-1)\geq 0$.  The operator $Q$ is not gauge invariant by definition, and may have negative anomalous dimension.   If $\mathcal{O}$ is a chiral operator, then its dimension is determined by its $R$-charge, dim$(\mathcal{O})=\frac{3}{2} R_\mathcal{O}$. So for chiral $\mathcal{O}$, $\mathcal{O}'$, we have dim$(\mathcal{O}\mathcal{O}') = \text{dim}(\mathcal{O}) + \text{dim}(\mathcal{O}')$.

If we have a theory with a Yukawa coupling of the form
\be
	y X_1 X_2 X_3
\ee
then the beta function for $y$ is
\be
	\beta_y = \frac{y}{2} (\gamma_1 + \gamma_2 + \gamma_3)
\ee
where $\gamma_i$ is the anomalous dimension of $X_i$. Because these $\gamma_i \geq 0$, the coupling flows to smaller values in the IR,
\be
	y(\mu) = y(\mu_0) \left( \frac{\mu}{\mu_0} \right)^{\frac{1}{2} (\gamma_1+\gamma_2+\gamma_3)}.
\ee

Including an additional weakly gauged $SU(3)\times SU(2) \times U(1)$ symmetry under which some $X_i$ and $Q_i$ may be charged, beta functions and anomalous dimensions should change by order $g_i^2/(16\pi^2)$, which is small if the new symmetry is weakly gauged, so if the anomalous dimensions were large without this new group, its addition should not qualitatively change any results. Compare this to the claim in section \ref{gaugesymmetries} that weakly gauged symmetries of a CFT correspond to bulk gauge symmetries.

This means if, in our beta function above, the $X_i$ correspond to Standard Model particles, and $\mu_0$ is the Planck scale, and the couplings involving $G$-charged fields $Q$ become conformal at a scale $M_c$, then our Yukawa couplings will in the IR run to
\be
	\label{ann:y}
	y \sim \left( \frac{M_c}{M_\text{P}} \right)^{\frac{1}{2} (\gamma_1+\gamma_2+\gamma_3)}.
\ee
This is the same result as was obtained in sections \ref{ren:fermions} and \ref{ren:scalar} for UV localized fields. The only difference here is that we have the additional weakly gauged Standard Model group making an additional order $g^2/16\pi^2$ contribution which we have argued is small enough to ignore. If the CFT causes the $X_i$ to develop large anomalous dimensions, then the coupling can be very small in the IR.

We can show that most terms that can appear in the superpotential correspond to irrelevant operators.  Consider a general-looking term in the superpotential,
\be
	W_c = c w.
\ee
The beta function for $c$ is
\be
	\beta_c = c (\text{dim } w-3),
\ee
but we know from before that $\text{dim }(cw) = \text{dim }c + \text{dim }w = \text{dim }W = 3$. The operator $w$ contains $k\geq 0$ powers of neutral fields $X$ times a $G$-invariant chiral $\mathcal{O}(Q)$, which means dim $w\geq k+1$ (unless $\mathcal{O}$ is the identity in which case $w\geq k$).  Then, if dim $w>3$ in the IR, this term flows to zero in the IR. This means the only marginal or relevant operators have the form
\begin{align}
	& \mathcal{O}(Q), \quad \text{dim }\mathcal{O}\leq 3 \\
	& X\mathcal{O}(Q), \quad \text{dim } \mathcal{O} < 2 \\
	& XX, \quad \text{dim }X \leq 3.
\end{align}

\subsection{General Predictions}

As a result of (\ref{ann:y}) we can write the Yukawa couplings in this theory in the form
\be
	Y_{ij} = \tilde{Y}_{ij} \epsilon_{L_i} \epsilon_{R_j}
\ee
where $\tilde{Y}$ is the short-distance coupling, and we have defined 
\be
	\epsilon_{L,R} = \left( \frac{M_c}{M_\text{P}}  \right)^{\gamma_{L,R}/2}.
\ee

From this, we can immediately make a number of order of magnitude estimates (see section 3 of \cite{Nelson:2000sn}).  The CKM mixing angles satisfy the order-of-magnitude relations
\begin{align}
	&\label{ckm1} V_{ud} \sim V_{cs} \sim V_{tb} \sim 1 \\
	& V_{ub} \sim V_{td} \sim V_{cb}V_{us}\\
	& V_{ts} \sim V_{cb}\\
	& \label{ckm4} V_{us} \sim V_{cd}
\end{align}
which are all satisfied in nature up to a factor of two.  Each $V_{ab}$ can be written in terms of the appropriate ratios of $\epsilon$s, but these relations are satisfied independently of the specific $\epsilon$s.

Because fermion masses are proportional to Yukawa couplings we also know,
\be
	\frac{m_i}{m_j} \sim \frac{\epsilon_{L_i} \epsilon_{R_i}}{\epsilon_{L_j} \epsilon_{R_j}}.
\ee

In section \ref{SMinbulk} we learned that the Yukawa couplings of the 4d theory could be written in terms of the 5d couplings as (\ref{y5d:general}),
\be
	\lambda_{ij} = \frac{\tilde{\lambda}_{ij} k }{N_i N_j} e^{(1-c_i-c_j)y^*}
\ee
with
\be
	N_i^{-2} = \frac{1/2-c_i}{e^{(1-2c_i)\pi k R}-1}.
\ee
We can symmetrically write this as
\be
	\lambda_{ij} = k \tilde{\lambda}_{ij} \left( \frac{e^{(\frac{1}{2}-c_i)y^*}}{N_i} \right)\left( \frac{e^{(\frac{1}{2}-c_j)y^*}}{N_j} \right)
\ee
\be
	=  \tilde{\lambda}_{ij} k \epsilon_i \epsilon_j.
\ee

Due to the dimensions of the 5d action, $\tilde{\lambda}k$ is dimensionless, and corresponds to the $\tilde{\lambda}$ in the previous section when units are chosen in which $k=1$. So the $\epsilon$ factors of \cite{Nelson:2000sn} which determine suppression are just our wave function localization terms. Mass ratios are also determined in terms of $\epsilon$s, so assuming the short distance couplings are all of order one, measuring fermion mass ratios and Yukawa couplings completely determines localization in the bulk.

As a result, because the relations (\ref{ckm1}-\ref{ckm4}) are satisfied \emph{independently} of the specific $\epsilon$s, and only depend on the form of the low-energy Yukawa couplings in terms of the short-distance ones, this form of the CKM is a generic prediction of Randall-Sundrum models with upgraded-to-5d Standard Model Yukawa couplings. A detailed discussion of CKM matrices in RSI can be found in \cite{Casagrande:2008hr}. We also note that these results are independent of the details of the particular CFT we have used.

\subsection{An SU(5) Model}

We can make additional predictions by assuming we have a specific GUT group, and although they are model-dependent we expect similar results to hold for different GUT groups.

If we assume that we have an $SU(5)$ group, and there are suppression factors for the three $\overline{5}$s, $\epsilon_{\overline{5}_r}$, and for the 10s, $\epsilon_{10_i}$, we can write the CKM matrix elements out explicitly in terms of them. 

The $SU(5)$ model predicts CKM elements \cite{Nelson:2000sn} 
\begin{align}
 	& V_{us} \sim V_{cd} \sim \frac{\epsilon_{10_1}}{\epsilon_{10_2}} \sim \sqrt{\frac{m_u}{m_c}} \\
	& V_{cb} \sim V_{ts} \sim \frac{\epsilon_{10_2}}{\epsilon_{10_3}} \sim \sqrt{\frac{m_c}{m_t}} 
\end{align}
and for leptons
\begin{align}
	& V_{e2} \sim \frac{\epsilon_{\overline{5}_1}}{\epsilon_{\overline{5}_2}} \sim \sqrt{\frac{m_{\nu_1}}{m_{\nu_2}}} \sim \frac{ m_e/m_\mu}{\sqrt{m_u/m_c}} \sim \frac{ m_d/m_s}{\sqrt{m_u/m_c}}\\
	& V_{\mu3} \sim \frac{\epsilon_{\overline{5}_2}}{\epsilon_{\overline{5}_3}} \sim \sqrt{\frac{m_{\nu_2}}{m_{\nu_3}}} \sim \frac{ m_\mu/m_\tau}{\sqrt{m_u/m_c}} \sim \frac{ m_s/m_b}{\sqrt{m_u/m_c}} .
\end{align}
These relations are satisfied to within an order of magnitude or so, but since, e.g., $m_d/m_s \sim 10m_e/m_\mu$, to get values for the ratios of epsilons, following the reference, we will take geometric means of the mass ratios to find
\begin{align}
	& \frac{\epsilon_{10_1}}{\epsilon_{10_2}} \sim 0.07, \quad 	\frac{\epsilon_{10_2}}{\epsilon_{10_3}} \sim 0.04 \\
	& \frac{\epsilon_{\overline{5}_1}}{\epsilon_{\overline{5}_2}} \sim 0.15, \quad 	\frac{\epsilon_{\overline{5}_2}}{\epsilon_{\overline{5}_3}} \sim 0.9.
\end{align}
From these, we can find ratios of wave function localization factors using
\be
	\epsilon_i = \frac{1}{N_i}  e^{(1/2-c_i )y^*} , \quad N_i^{-2} = \frac{1/2-c_i}{e^{(1-2c_i)\pi k R}}.
\ee
There are exact solutions for the $c_{i,j}$s in $\epsilon_i/\epsilon_j = \#$ in terms of special functions, but we can numerically solve this to see what values of $c$ are allowable.  We find the dependence on the specific value of the ratio is not strong, so we take $\epsilon_i/\epsilon_j = 0.01,$ $0.1$, and $0.2$. 

\begin{figure}[h]
	\centering
	\subfloat[IR Higgs]{\includegraphics[width=7cm, height=7cm]{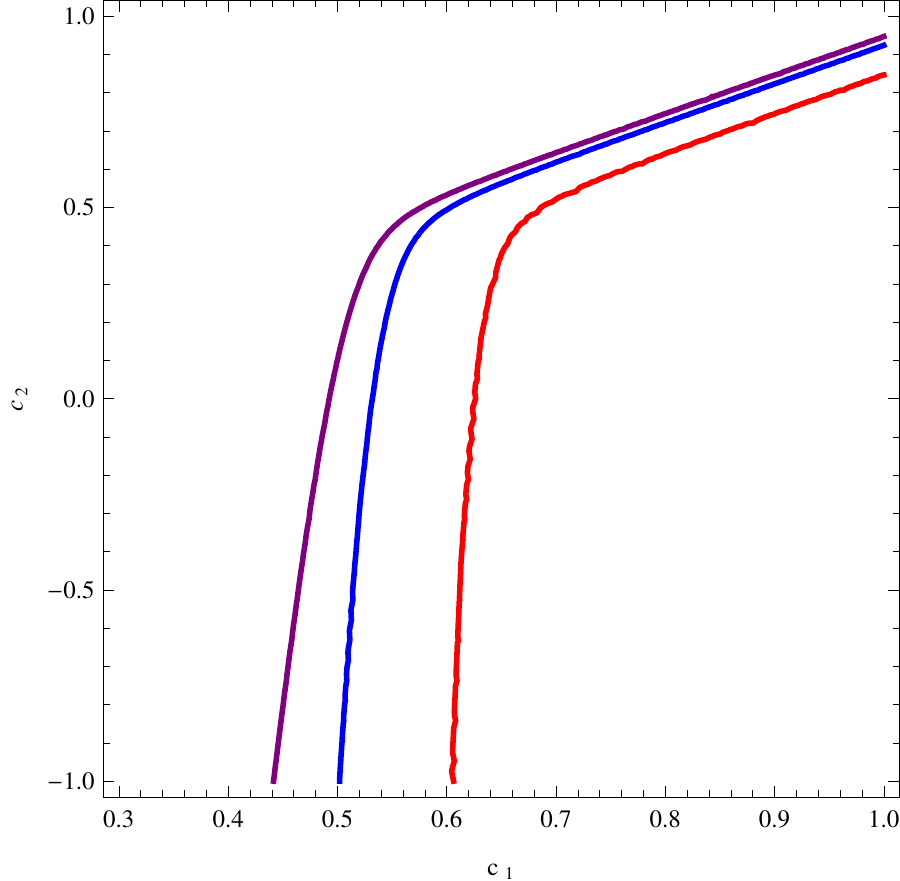}}                
	\subfloat[UV Higgs]{\includegraphics[width=7cm, height=7cm]{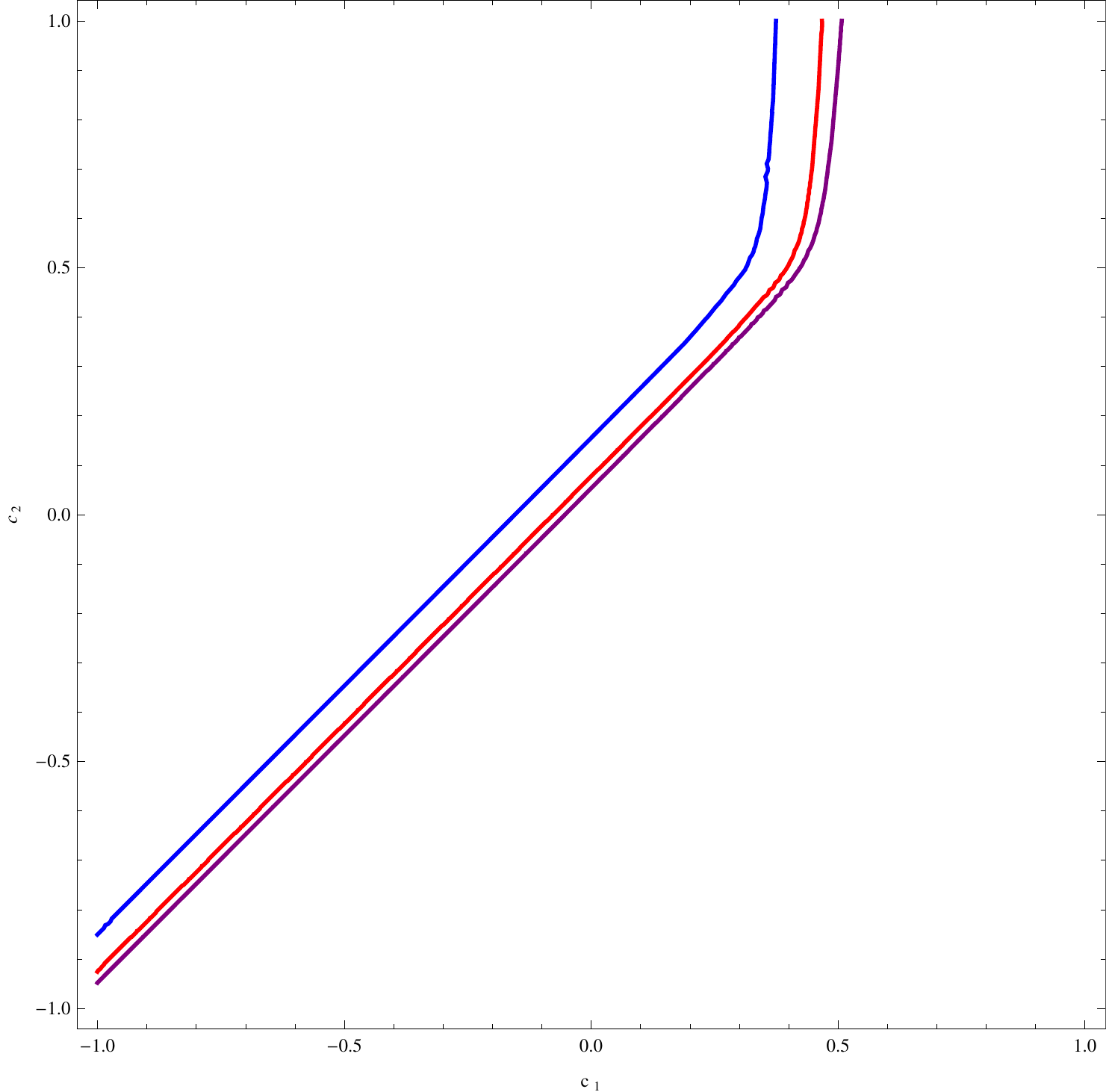}}
	\caption{The right/center line corresponds to $0.01$, the center/left to $0.1$, and the left/right to $0.2$. For the purposes of the graph, we have chosen $kR=10$, but the shape of the graph is not dependent on this.  Each axis corresponds to IR (left/bottom) to UV (right/top) localization.}
\end{figure}

For an IR-localized Higgs, $y^*=\pi R$, we see that for one fermion localized anywhere in the bulk, in order to get these specific order of magnitude relations, the other is forced away from the IR. The second fermion can be either weakly IR localized, flat, or UV localized. This means all fermions can not be strongly IR localized and be consistent with this setup.

The opposite conclusion is reached for a UV-localized Higgs, $y^* = 0$, where one fermion can be anywhere, but the other cannot be too UV-localized. In this case, it appears much more difficult to localize the first fermion to be near the Higgs than in the previous case. Here, fermions cannot all be UV localized.

We can say a little more about localization by noting that in the setup to this, we have assumed that the Higgs is a Standard Model field, and not a CFT composite. This means that from the point of view of the 5d theory, the Higgs is UV-localized (or lives on the UV brane), so the IR-localized fermions will have the smallest Yukawa couplings, and there should be a generational ordering of localizations,
\be
	\text{IR} < c_\text{1st generation} < c_\text{2nd generation} < c_\text{3rd generation} < \text{UV}.
\ee
This generational ordering corresponds to an ordering of the basis of the CKM matrix by masses noted in \cite[section 3]{Nelson:2000sn}, which is automatically obtained by writing it in the basis it is most diagonal in.

None of the results of this section depend on the specific CFT involved. The only assumption made was the UV-localization of the Higgs, required in order to make it an elementary field.  This assumption is not required, and models with CFT composite Higgses exist, but making this assumption will change some of qualitative conclusions of this section; particularly, it will reverse the localization of fermions. The same qualitative conclusions also hold for groups other than $SU(5)$.

\subsection{Anomalous Dimensions}

One difficult aspect of these 4d models is that, while the dimension of chiral operators is determined by the $R$ charge,
\be
	\text{dim }\mathcal{O} = \frac{3}{2} R_\mathcal{O}
\ee
it may be difficult to find this symmetry exactly in the SCFT, to find the $R$ charge.  However, we know from earlier that the operator dimension is determined by localization. For example, in the case of a fermion in the bulk,
\be
	\text{dim }\mathcal{O}_\psi = \frac{3}{2} + \left| c \pm \frac{1}{2} \right| = \frac{5}{2} + \gamma
\ee
where 
\be
	\gamma = \alpha - 1.
\ee
Operator dimensions are very easy to calculate from the 5d dual.

In the stringy realizations of models in a slice of AdS, such as the Klebanov-Strassler solution, the full theory is defined on manifolds like $AdS_5\times T^{1,1}$, and the $R$ symmetry of the CFT corresponds to cycles of $T^{1,1}$.  In this case fluxes through the bulk, which know about the $T^{1,1}$, close off the IR end and create the IR brane, causing the CFT dual to break down at some scale.  

In the RSI case there is no corresponding extra dimension for this symmetry, but the localization of fields with respect to the IR brane encodes the same information.

We can easily see from the definition of the anomalous dimension in terms of bulk mass parameters that a large anomalous dimension, which is needed in the 4d theory to develop small couplings, corresponds to some fields having an extreme UV or IR localization. For example, for the $+$ branch in the fermion case above, $c\gg \frac{1}{2}$ or $c\ll \frac{1}{2}$ correspond to large $\gamma$.

\subsection{Conclusions}

The 4d model has several general features:
\begin{itemize}
	\item {\bfseries CKM matrix element relations.} This model predicts that the order of magnitude form of CKM matrix elements has the same form as the Wolfenstein parameterization, and makes additional predictions for mixing angles for leptons, and neutrino oscillations \cite[section 3]{Nelson:2000sn}. These do not depend on any assumptions, other than that the coupling to some unspecified CFT drives anomalous dimensions to be large. The only requirement is the form of the Yukawa couplings, which follows from the coupling of some fields to a CFT, which causes large anomalous dimensions to develop.
	\item {\bfseries GUT-type relations.} If specific anomalous dimensions are not known, but specific suppression factors are known to come from specific GUT multiplets, then ratios of suppression factors can be related to CKM matrix elements. 
	\item {\bfseries Anomalous dimensions.} SCFTs determine operator dimensions of chiral operators in terms of $R$ charge through dim $\mathcal{O}=\frac{3}{2}R_\mathcal{O}$. If this symmetry in the SCFT is known explicitly, $R$ charges, and anomalous dimensions can be calculated. Large anomalous dimensions are required to generate small couplings.
	\item {\bfseries Generational ordering.} Ordering the CKM matrix in the basis where it is the most diagonal automatically corresponds to ordering by mass.
\end{itemize}

Each of these general features corresponds to elements of the 5d dual:
\begin{itemize}
	\item {\bfseries CKM matrix element relations.} CKM matrix elements are determined in the by Yukawa matrices, which in the bulk are entirely determined by localization of fields. Surprisingly, this makes specific order of magnitude predictions for arbitrary localizations. 
	\item {\bfseries GUT-type relations.} Knowing specific fields in the bulk (corresponding to knowing specific GUT representations in the 4d theory) automatically determine ratios of suppression factors, because their wave function profiles are known.
	\item {\bfseries Anomalous dimensions.} The large anomalous dimensions required for suppression factors in the 4d dual are generated by extreme localization of some of the bulk Standard Model fermions. 
	\item {\bfseries Generational ordering.}  Generational ordering is the result of the ordering of fermions in the bulk, which are ordered by their bulk masses, which determines their Yukawa couplings calculated from their overlap with the Higgs, which determine masses. 
\end{itemize}

This correspondence can be made explicit through the introduction of two relations:
\begin{align}
	& \epsilon_\text{4d} = \left(\frac{M_\text{CFT}}{M_P}\right)^{\frac{\gamma}{2}} \nonumber\\
	& \quad\quad\quad = \\
	& \epsilon_{5d} = \frac{e^{(\frac{1}{2} - c_i)}y^*}{N}. \nonumber
\end{align}
\begin{align}
	& \text{dim }\mathcal{O}_\psi = \frac{3}{2}R_\mathcal{O} \nonumber\\
	& \quad\quad\quad = \\
	& \text{dim }\mathcal{O}_\psi = \frac{3}{2} + \left| c + \frac{1}{2} \right|
\end{align}
which tells us that the general predictions of each model are very simply encoded in just a few parameters.

Additionally, the same large anomalous dimensions that drive the Yukawa couplings to small values can drive dangerous FCNC and EDM terms to zero \cite{Nelson:2000sn}. This does not depend on any new features of the models other than the specific forms of the couplings. This corresponds to the claims in the very beginning of these notes, that a promising possibility was that the exponential warp factor could suppress terms like these.

In this section we have seen that many of the general features found in a four dimensional model proposed by thinking about renormalization group flow can be reinterpreted as coming from the bulk structure of an entirely different looking five dimensional model.  We've shown that the five dimensional models inherit the promising phenomenological features of the four dimensional model, and have explicitly constructed their bulk counterparts.  It is hoped that further investigation of general features of models of this sort can shed some light on additional features of the Standard Model that are not well understood.

\section{Conclusions}

Holography provides two interesting alternative way to understand extensions of the Standard Model. They can be understood from the low energy point of view, by writing down a phenomenologically motivated 4d model, such as in section 11, and trying to understand the properties of the 5-dimensional dual. Theories can also be understood from a high energy point of view, by writing down a 5d gravitational theory, and finding its 4d SCFT dual.

These models can make distinctive and definite predictions, such as the structure of CKM matrices, generational couplings, flavor universality, and specrta of superpartners and Kaluza-Klein states, that could be visible at the LHC if the IR scale is low enough. Even if it is not low enough to see directly, the general predictions that can be made by these models give promising explanations of features of the Standard Model that are not well understood.

\section*{Acknowledgments}
\addcontentsline{toc}{section}{Acknowledgments}

I would like to thank my thesis adviser Ann Nelson for showing me how interesting the world of phenomenology is, for always giving me new things to think about, and for sharing her insight and experience with me.

I would also like to thank Andreas Karch, for always being willing to talk with me about whatever I've been trying to understand, and for organizing the weekly AdS/CFT lunch which always provides interesting topics to think about.

Finally, I would like to thank the graduate students and postdocs here that have provided me with many interesting and insightful discussions.

\bibliographystyle{utphys}
\bibliography{bibliography}

\end{document}